\documentstyle[prb,aps,psfig,floats]{revtex}
\begin{document}
\author{A.A.Ovchinnikov, V.Ya.Krivnov, and D.V.Dmitriev}
\address{Joint Institute of Chemical Physics of RAS, 117977 Moscow, Russia\\
Max-Planck-Institut fur Physik Komplexer Systeme, 01187 Dresden, Germany \\} 
\title{Exact ground state of one- and two-dimensional 
frustrated quantum spin systems.}
\maketitle
\begin{abstract}
We outline the recent results on the ground state for a class of one- 
and two-dimensional frustrated quantum spin models with competing ferro(F)- 
and antiferromagnetic (AF) interactions. 
Frustrated spin systems are known to have many interesting properties due
to large quantum fluctuations. As a result of these fluctuations usual
mean-field approach gives quite crude (if not false) description of these
systems. Therefore, exactly solvable models are very instructive at
investigations of such systems.
The exact ground state wave function of proposed models
has a structure of the valence-bond state (VBS) type.
One of the 1D model describes the transition line between the F and AF
phase. The exact singlet ground state on this line has a double-spiral
ordering. Using different approximation methods we study the magnetization
curve in the AF phase.
The second considered set of the 1D and 2D models has an exact
non-degenerate ground state with exponentially decaying spin correlations.
We also proposed the 1D and 2D electronic models with exact ground state
represented in terms of singlet bond functions which are the generalization
of the RVB functions including ionic states.
\end{abstract}


\section{Introduction}

There is currently much interest in quantum spin systems that exhibit
frustration \cite{1}. This has been stimulated in particular by the study of
the magnetic properties of the cuprates which become high-Tc superconductors 
when doped. Frustrated spin systems are known to have many interesting 
properties that are quite unlike those of conventional magnetic systems. 
The simplest model of such kind is the Heisenberg spin chain with nearest- 
and next-nearest neighbor interactions $J_1$ and $J_2$. This model is well 
studied for $J_1,\,J_2>0$ \cite{45,3,7}. In particular, it has been found 
that at $J_2=0.24J_1$ the transition from the gapless state to the dimerized 
one takes place \cite{8}. The point $J_2=J_1/2$ corresponds to the well-known 
Majumdar-Ghosh model \cite{39} for which the exact ground state consists of 
dimerized singlets and there is a gap in the spectrum of excited states. 

Less studied is frustrated spin models with competing interactions of ferro- 
and antiferromagnetic types. The physical interest for these models is
connected with the study of the real compounds containing $CuO$ chains
with edge-sharing $CuO_4$ units, like $La_6Ca_8 Cu_{24}O_{41}$, $Li_2CuO_2$
and $Ca_2Y_2Cu_5O_{10}$ \cite{i1}. In these compounds $Cu-O-Cu$ bond angle is
nearly $90^{\circ}$ and the nearest-neighbor $Cu-Cu$ spin interaction is 
ferromagnetic according to Goodenough-Kanamori-Anderson rule \cite{i2} while 
the next-nearest-neighbor interaction is antiferromagnetic.
The magnetic properties of these models are very
different from those for models with pure antiferromagnetic interactions.
Their ground state can be either ferromagnetic or singlet depending on the 
relation between the exchange integrals. One of the most interesting problems 
related to these models is the character of the transition between 
ferromagnetic (F) and antiferromagnetic (AF) phases.

Of a special importance are models for which it is possible to construct 
an exact ground state. Last years a considerable progress has been achieved 
by using so-called Matrix-Product (MP) form of the ground state wave function 
\cite{27,13}. The ground state wave function in the MP method is represented 
by Trace of a product of matrices describing single-site states. 
The MP ground state has a structure of the type where each neighboring pair 
of spins has valence bond and, in fact, the MP form is a convenient 
representation of the valence bond states. Its origin can be traced to the
S=1 quantum spin chain with bilinear and biquadratic interactions \cite{40}.
At present, various 1D spin models with the exact MP ground state is found
\cite{13,prb,25}.

In this paper we present a set of 1D and 2D spin-1/2 models with competing F 
and AF interactions for which the singlet ground state wave function can be 
found exactly. This function has a special form expressed in terms of 
auxiliary Bose operators. This form of the wave function is similar to the
MP one but with infinite matrices. For special values of model parameters it
can be reduced to the standard MP form. 

One of models is 1D quantum spin model describing the F-AF
transition point. Spin correlations in the singlet ground state show giant 
spiral magnetic structure with period equals to the system size. On the 
antiferromagnetic side of this point the ground state can be either gapless 
with incommensurate spiral ground state \cite{ko} or gapped with exponential 
decay of correlators \cite{epj}. There are regions in the AF phase where the 
magnetization as a function of magnetic field has jumps.

The second considered model is the special case of the spin ladder with 
exchange integrals depending on one parameter. The exact ground state of 
this model is non-degenerated singlet with exponentially decaying spin 
correlations, and there is an energy gap.

It will be shown that proposed form of exact wave function can be generalized 
to higher dimensions and two dimensionalal frustrated spin model with exact 
ground state can be constructed.

It is known that the exact ground state of some 1D and 2D quantum spin models
can be represented in RVB form \cite{42,39,44,40,jetp}. The RVB 
function in the Fermi representation consists of homopolar configurations 
of electron pair. We generalize this two-particle function to include ionic 
states and denote it as "singlet bond" (SB) function. It is
natural to try to find electronic models with exact ground state formed by SB
functions in the same manner as for known spin models. In the paper some models
of interacting electrons are presented. The Hamiltonians of these models 
include the correlated hopping of electrons and spin interactions. One of the 
1D models describes the transition point between the phases with and without 
an off-diagonal long-range order.

The paper is organized as follows. In Sec.2 we consider the frustrated spin 
chain at F-AF transition point and describe the exact singlet ground state 
wave function as well as details of the spin correlation function 
calculations. We discuss the phase diagram of this model and its magnetic 
properties in the AF phase. In Sec.3 the special spin ladder will be 
considered. Two-dimensional frustrated spin model with the exact ground state 
is considered in Sec.4. Sec.5 is devoted to the construction of the electronic
models with SB type of wave function. The results of paper is summurized in 
Sec.6.

\section{Zigzag spin model.}

\subsection{Zigzag spin model at F-AF transition point.}

Let us consider $s=\frac 12$ spin chain with nearest- and next-nearest
neighbor interactions given by the Hamiltonian
\begin{equation}
H =   -\sum_{i=1}^M({\bf S}_{2i-1}\cdot{\bf S}_{2i}-\frac 14)
+J_{23}\sum_{i=1}^M({\bf S}_{2i}\cdot{\bf S}_{2i+1}-\frac 14)
+J_{13}\sum_{i=1}^N({\bf S}_i\cdot{\bf S}_{i+2}-\frac 14) ,
\label{a1}
\end{equation}
with periodic boundary conditions and even number of spins $N=2M$.
This model is equivalent to a ladder model with diagonal 
coupling, so called a `zigzag' spin ladder.

If $J_{23}<1$, then the ground state of (\ref{a1}) is ferromagnetic 
(singlet) at $\delta <0$ ($\delta >0$), where 
$\delta =J_{23}+\frac{2J_{13}}{1-2J_{13}}$ (Fig.\ref{phase}). 
The equation $\delta =0$ defines the line of transition points from
the ferromagnetic to the singlet state, when energies of these states are
zero. The model (\ref{a1}) along this line is given by the Hamiltonian 
depending on the parameter $x$ ($x>-1/2$):
\begin{figure}[t]
\unitlength1cm
\begin{picture}(8,5)
\centerline{\psfig{file=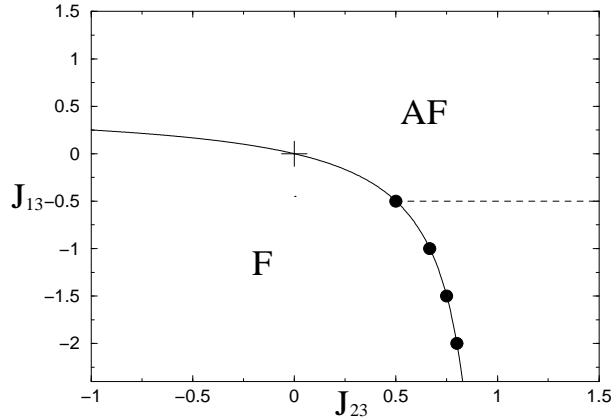,angle=0,width=8cm}}
\end{picture}
\vspace{1cm}
\caption[]{Phase diagram of the `zigzag' model (\ref{a1}). 
The solid line is the 
boundary between the ferromagnetic and singlet phases. Circles
correspond to the special cases of the model. 
On the dotted line the ground state is a product of singlet pairs.
\label{phase} }
\end{figure}
\begin{equation}
H = -\sum_{i=1}^M({\bf S}_{2i-1}\cdot{\bf S}_{2i}-\frac 14)
  + \frac{2x}{2x+1}\sum_{i=1}^M({\bf S}_{2i}\cdot{\bf S}_{2i+1}-\frac 14)
  - x\sum_{i=1}^N({\bf S}_i\cdot{\bf S}_{i+2}-\frac 14) , 
\label{a2}
\end{equation}
with periodic boundary conditions.

We note that the Hamiltonian (\ref{a2}) has a symmetry: 
its spectrum coincides with the spectrum of $\widetilde{H}(x)$ 
obtained by the following transformation
\[
\widetilde{H}(x)=-\frac{2x}{2x+1}\; H(-x-\frac 12),\qquad -\frac 12<x<0
\]

This transformation permutes the factors at the first and the second terms
in the Hamiltonian (\ref{a2}). Thus, due to the symmetry it is sufficient to
consider the range $x \geq -\frac 14$.

First, we will show that the ground state energy of (\ref{a2}) is zero. Let us
represent the Hamiltonian (\ref{a2}) as a sum of Hamiltonians $h_n$ of cells
containing three sites
\begin{equation}
H=\sum_{i=1}^M(h_{2i-1}+h_{2i}),  
\label{a3}
\end{equation}
where
\[
h_{2i-1}= -\frac 12({\bf S}_{2i-1}\cdot{\bf S}_{2i}-\frac 14)
          +\frac{x}{2x+1}({\bf S}_{2i}\cdot{\bf S}_{2i+1}-\frac 14)
          -x ({\bf S}_{2i-1}\cdot{\bf S}_{2i+1}-\frac 14) ,
\]
\[
h_{2i}= -\frac 12({\bf S}_{2i+1}\cdot{\bf S}_{2i+2}-\frac 14)
        +\frac{x}{2x+1}({\bf S}_{2i}\cdot{\bf S}_{2i+1}-\frac 14)
        -x ({\bf S}_{2i}\cdot{\bf S}_{2i+2}-\frac 14)
\]
Eigenvalues of each $h_n$ are
\[
\lambda_1=\lambda_2=0,\qquad \lambda_3=\frac{4x^2+2x+1}{4x+2}>0
\]

We will present a singlet wave function which is the exact one of each $h_n$
with zero energy and, therefore, it is the exact ground state wave function
of (\ref{a2}). This function has a form
\begin{equation}
\Psi_0 = P_0 \Psi , \qquad
\Psi = \langle 0_b|\, g_1\otimes g'_2\otimes g_3\otimes\ldots\otimes g'_N \, 
|0_b \rangle  
\label{a4}
\end{equation}
where
\begin{equation}
g_i = b^+\left|\uparrow\right\rangle_i 
        +\left|\downarrow\right\rangle_i , \qquad
g'_i = (b^+b-2x)\left|\uparrow\right\rangle_i 
             + b\left|\downarrow\right\rangle_i
\label{a5}
\end{equation}
Here we introduced one auxiliary Bose-particle $b^+$ (the Bose operators
$b^{+}$ and $b$ do not act on spin states $\left|\uparrow\right\rangle _i$ 
and $\left|\downarrow\right\rangle_i$) and the Bose vacuum 
$\left|0_b\right\rangle$. Therefore, the direct product 
$g_1\otimes g'_2\otimes\ldots\otimes g'_N$ is the superposition of all 
possible spin configurations multiplied on the corresponding Bose
operators, like $b^{+}b\,b\,b^{+} \ldots \left|\uparrow\downarrow\downarrow
\uparrow \ldots\right\rangle$. $P_0$ is a projector onto the singlet state.
This operator can be written as \cite{18}
\begin{equation}
P_0=\frac 1{8\pi ^2}\int_0^{2\pi }d\alpha \int_0^{2\pi }d\beta \int_0^\pi
\sin \gamma d\gamma ~e^{i\alpha S^z}e^{i\gamma S^x}e^{i\beta S^z},
\label{a6}
\end{equation}
where $S^{x(z)}$ are components of the total spin operator.

The form of wave function (\ref{a4}) resembles the MP form, but with an 
infinite matrix
which is represented by Bose operators. Therefore, we have to pick out the
$\langle 0_b|\ldots |0_b \rangle $ element of the matrix product instead of 
the usual trace in the MP formalism \cite{27,13}, because 
the trace is undefined in this case. 
The function $\Psi $ contains components with all possible values of 
spin $S$ ($0\leq S\leq N/2$) and, in fact, a fraction of the singlet is
exponentially small at large $N$. This component is filtered out by the
operator $P_0$.

One can easily check that each cell Hamiltonian $h_{2i-1}$ and $h_{2i}$ 
for $i=1,\ldots (M-1)$ gives zero when acting on the corresponding part 
in wave function $\Psi$ -- $g_{2i-1}\otimes g'_{2i}\otimes g_{2i+1}$ and
$g'_{2i}\otimes g_{2i+1}\otimes g'_{2i+2}$
(one should take care of Bose commutation relations).
Since each $h_i$ is a non-negatively defined operator, then $\Psi$ is 
the exact ground state wave function of an open chain:
\begin{equation}
H_{\rm open}=\sum_{i=1}^{M-1} (h_{2i-1}+h_{2i})
\label{a7}
\end{equation}

As mentioned above, the function $\Psi$ contains components 
of all possible values of total spin $S$, and, therefore, the ground 
state of the open chain is multiply degenerate. However it can be proven
\cite{zp,prb} that for the cyclic chain (\ref{a3}) only singlet and 
ferromagnetic components of $\Psi $ have zero energy. Therefore, for cyclic 
chain (\ref{a3}) $\Psi_0$ is the singlet ground state wave 
function degenerate with the ferromagnetic state.

In particular case, $x =-1/4$, when $J_{12}=J_{23}=-1$ and $J_{13}=\frac 14$,
another form of the exact singlet ground state wave function has been found
in \cite{42}. It reads
\begin{equation}
\Psi =\sum [i,j][k,l][m,n]\ldots ,
\label{a8}
\end{equation}
where $[i,j]$ denotes the singlet pair and the summation is made for any
combination of spin sites under the condition that $i<j,k<l,m<n\ldots$.

The following general statements relevant to the Hamiltonian (\ref{a2}) 
can be proved:

1). The ground states of open chain described by (\ref{a7}) in the sector 
with fixed total spin $S$ are non-degenerate and their energies are zero.

2). For cyclic chains the ground state in the $S=0$ sector is non-degenerate. 
The ground state energies for $0<S<M$ are non-zero.

3). The singlet ground state wave function for open and cyclic chains
coincide with each other.

\subsection{Spin correlations in the ground state.}

For the sake of simplicity we show the calculation of the spin correlation
function in the symmetric case $x=-1/4$, when the Hamiltonian (\ref{a2})
takes the form
\begin{equation}
H  =   - \sum_{i=1}^N ({\bf S}_i\cdot {\bf S}_{i+1}-\frac 14)
+\frac 14\sum_{i=1}^N ({\bf S}_i\cdot {\bf S}_{i+2}-\frac 14),
\label{f1}
\end{equation}

Since in this case there is
one spin in elementary cell, the singlet ground state wave function $\Psi_0$
can be written in a more simple and symmetric form:
\begin{equation}
\Psi_0 = P_0 \Psi , \qquad
\Psi = \langle 0_b|\, g_1\otimes g_2\otimes\ldots\otimes g_N\, |0_b\rangle , 
\label{b1}
\end{equation}
where
\begin{equation}
g_i = b^{+}\,\left|\uparrow\right\rangle _i 
      + b \, \left|\downarrow\right\rangle _i
\label{b2}
\end{equation}

One can check that wave functions (\ref{b1}) is the singlet ground state 
wave function with zero energy of Hamiltonian (\ref{f1}). Therefore, 
the equivalence of wave functions (\ref{a4}) and (\ref{b1}) follows from 
the non-degeneracy of ground state in the $S=0$ sector (though functions
$\Psi$ in Eqs.(\ref{b1}) and (\ref{a4}) are different).

Now we calculate the norm and correlation function of the wave function
$\Psi_0$ (\ref{b1}). The norm of the singlet wave function $\Psi_0$ is
\begin{equation}
\langle \Psi _0| \Psi _0\rangle =\langle \Psi |\: P_0\:| \Psi \rangle
\label{b3}
\end{equation}
It is easy to check that the function $\Psi $ has $S^z=0$. Then the 
projector $P_0$ in Eq.(\ref{a6}) takes the form \cite{prb}
\begin{equation}
P_0=\frac 12\int_0^\pi \sin \gamma d\gamma \,  
e^{izS^{-}} e^{iz^{\prime }S^{+}} ,
\label{b4}
\end{equation}
where $z=\tan \frac{\gamma }{2},\; z^{\prime }=\sin \frac{\gamma }{2} 
\cos \frac{\gamma }{2}$ and $S^{+(-)}$ are the operators of the total spin.

Therefore, the norm takes the form
\[
\langle \Psi _0| \Psi _0\rangle = \frac 12 \int_0^\pi \sin \gamma d\gamma \,
\langle 0_a , 0_b | \prod_{i=1}^{N} (g_{i}^{+} \, e^{izS_i^{-}} e^{iz^{\prime
}S_i^{+}} g_{i}) |0_a , 0_b  \rangle ,
\]
where
\begin{eqnarray*}
g_{i}^{+} \, e^{izS_i^{-}} e^{iz^{\prime }S_i^{+}} g_{i}
&=& (a^{+} \left\langle\uparrow _i\right| + a \left\langle\downarrow _i\right|
\,) \, e^{izS_i^{-}} e^{iz^{\prime }S_i^{+}}
(b^{+}\left|\uparrow\right\rangle _i +b\left|\downarrow\right\rangle _i \,)  \\
&=& a^{+}b^{+}+(1-z^{\prime }z)ab+izab^{+}+iz^{\prime }a^{+}b ,
\end{eqnarray*}
and $a^{+}$ and $a$ are the Bose operators. Thus the norm can be rewritten as 
\begin{equation}
\langle \Psi _0 | \Psi _0\rangle =\frac 12 \int_0^\pi \sin \gamma d\gamma \,
\langle 0| G^N |0\rangle ,  
\label{b5}
\end{equation}
where $\left|0\right\rangle=\left|0_a ,0_b\right\rangle$ is the Bose vacuum of
$a^{+}$ and $b^{+}$ particles and
\[
G=u(a^{+}b^{+}+ab)+iv(ab^{+}+a^{+}b) ,
\]
where $u=\cos \frac{\gamma}{2}, \; v=\sin \frac{\gamma}{2}$.

Let us introduce the auxiliary function $P(\xi )$:
\begin{equation}
P(\xi )=\langle 0| e^{\xi G} |0\rangle ,
\label{b6}
\end{equation}
then 
\[
\langle 0| G^N |0\rangle =\left. \frac {d^N P}{d\xi ^N} \right|_{\xi =0}
\]
The function $P(\xi )$ can be easily found \cite{ele}:
%
%
\begin{equation}
P(\xi)=\frac{1}{\sqrt{1-u^2 \sin^2 \xi }}
\label{b11}
\end{equation}
Integrating Eq.(\ref{b11}) over $\gamma $, we obtain
\begin{eqnarray}
\langle\Psi_0|\Psi_0\rangle =\left.\frac 12\int_0^\pi\sin\gamma d\gamma \;
\frac {d^N P}{d\xi ^N} \right|_{\xi =0}
=\left. \frac {d^N}{d\xi ^N} \left( \frac {1}{\cos^2(\frac{\xi }{2})} 
\right)  \right|_{\xi =0}
\label{b12}
\end{eqnarray}
Thus, finally, we arrive at
\begin{equation}
\langle \Psi _0| \Psi _0\rangle =
\left. 2 \frac {d^{N+1}}{d\xi ^{N+1}} \left( \tan \frac{\xi}{2}  \right)  
\right|_{\xi=0}= \frac{4\,(2^{N+2}-1)}{N+2}|B_{N+2}|
\label{b13}
\end{equation}
Here $B_N$ are the Bernoulli numbers.

To calculate the spin correlators we need to introduce operators
\begin{eqnarray}
G_z &=& g_{i}^{+} \, e^{izS_i^{-}} e^{iz^{\prime }S_i^{+}}2S_i^z g_{i}
= u(a^{+}b^{+}-ab)+iv(ab^{+}-a^{+}b) , \nonumber \\
G_{+} &=& g_{i}^{+} \, e^{izS_i^{-}} e^{iz^{\prime }S_i^{+}}S_i^{+} g_{i}
= ua^{+}b+ivab , \nonumber \\
G_{-} &=& g_{i}^{+} \, e^{izS_i^{-}} e^{iz^{\prime }S_i^{+}}S_i^{-} g_{i}
= uab^{+}+iva^{+}b^{+} \nonumber
\end{eqnarray}
Then, the correlator $\langle {\bf S}_1\cdot {\bf S}_{l+1}\rangle$ 
will be defined by
\begin{equation}
\langle \Psi _0 |{\bf S}_1\cdot {\bf S}_{l+1}|\Psi _0\rangle 
=\frac 12 \int_0^\pi \sin \gamma d\gamma \,
\langle 0|\: \frac 14 G_z G^l G_z G^{N-l-2}+
\frac 12 G_{+} G^l G_{-} G^{N-l-2}  |0\rangle   
\label{b14}
\end{equation}
(since $\langle 0|G_{-}\ldots |0\rangle =0 $).

The expectation values in Eq.(\ref{b14}) can be represented as
\begin{eqnarray}
\langle 0|\: G_z G^l G_z G^{N-l-2} |0\rangle &=& 
\frac {\partial ^l}{\partial \xi ^l} \frac {\partial ^{N-l-2}}
{\partial \zeta ^{N-l-2}} 
\left. \langle 0|G_z e^{\xi G} G_z e^{\zeta G} |0\rangle
\right|_{\xi =\zeta =0}  , \nonumber \\
\langle 0|\: G_{+} G^l G_{-} G^{N-l-2} |0\rangle &=& 
\frac {\partial ^l}{\partial \xi ^l} \frac {\partial ^{N-l-2}}
{\partial \zeta ^{N-l-2}} 
\left. \langle 0|G_{+} e^{\xi G} G_{-} e^{\zeta G} |0\rangle
\right|_{\xi =\zeta =0}
\label{b15}
\end{eqnarray}
After a procedure similar to that for the norm and the integration 
over $\gamma $, we obtain
\begin{equation}
\langle \Psi _0 |{\bf S}_1\cdot {\bf S}_{l+1}| \Psi _0\rangle 
= \frac {\partial ^l}{\partial \xi ^l} \frac {\partial ^{N-l-2}}
{\partial \zeta ^{N-l-2}} \left. \left( -\frac 38 \frac 
{\cos (\xi -\zeta )}{\cos^4(\frac{\xi +\zeta }{2})} \right)  
\right|_{\xi =\zeta =0}
\label{b16}
\end{equation}

It can be shown that in the thermodynamic limit, Eqs.(\ref{b13}) and
(\ref{b16}) result in
\begin{equation}
\left\langle {\bf S}_{i}\cdot {\bf S}_{i+l}\right\rangle =
\frac 14\cos \left( \frac{2\pi l}{N}\right)   
\label{b17} 
\end{equation}
So, we reproduce the result obtained in \cite{42,prb} that
in the thermodynamic limit a giant spiral spin structure is realized, 
with the period of the spiral equal to the system size.

For the general case of model (\ref{a2}) the calculation of the singlet 
ground state correlation functions can be performed by the similar way. 
The final result in the thermodynamic limit is \cite{zp,prb}:
\begin{eqnarray}
\left\langle {\bf S}_{n}{\bf S}_{n+2l}\right\rangle &=&
\frac 14 \cos \left( \frac{4\pi l}{N}\right), \nonumber \\
\left\langle {\bf S}_{n}{\bf S}_{n+2l-1}\right\rangle &=&
\frac 14 \cos \left( \frac{2(2l-1)\pi}{N} +(-1)^n \triangle \varphi \right) 
\label{b18}
\end{eqnarray}

The latter equations mean that the long-range double-spiral order exists
in the singlet ground state of Hamiltonian (\ref{a2}). The pitch
angle of each spiral is $\frac{4\pi}{N}$ and there is a small shift angle 
$\triangle\varphi =\frac{2\pi (4x+1)}{N}$ between them.
This shift angle reflects the fact that the unit cell contains two sites
unless $x=-\frac 14$.

We note that for special values of model parameter $x=\frac 12,1,\frac 32,2
\ldots$ Eqs.(\ref{b18}) are not valid and spin correlations decay 
exponentially. These cases will be considered in Sec.III (see Eqs.(\ref{d4})).

\subsection{Phase diagram of `zigzag' model}

\begin{figure}[tbp]
\unitlength1cm
\begin{picture}(13,9)
\centerline{\psfig{file=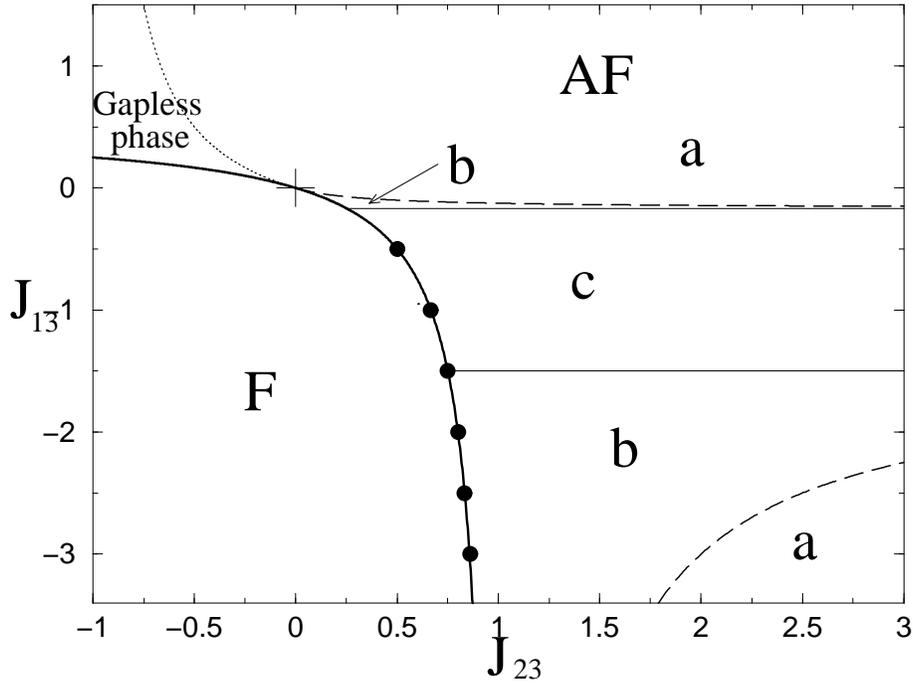,angle=0,width=12cm}}
\end{picture}
\vspace{5mm}
\caption{ \label{ph2} The thick solid line is the 
boundary between the ferromagnetic and singlet phases. Circles
correspond to the special cases of the model. The dotted line
denotes the heuristic boundary between gapped and gapless phases.
The dashed line is the boundary of the region with multimagnon
bound state.}
\end{figure}

So far we considered the model (\ref{a1}) at the transition line from 
ferromagnetic to antiferromagnetic state. Now we discuss the phase diagram 
of this model. The exact ground state in the AF phase is generally unknown. 
But it is interesting to note that the ground state on the line $J_{13}=-1/2$ 
is the product of singlets on ladder diagonals (2,3), (4,5), ... as in the
point $J_{13}=-J_{23}=-1/2$ on the transition line. The spectrum of (\ref{a1})
on the transition line is gapless. There are some regions on the plane 
($J_{13}$, $J_{23}$) which were studied by different approximation methods.

First let us consider behaviour of the system in the transition region
near the `symmetric' point $x=-\frac 14$, where
\begin{equation}
J_{12}=J_{23}=-1,\qquad J_{13}=\frac 14 +\delta
\label{b19}
\end{equation}

It would be mentioned that several copper oxide compounds are described 
by this model. These compounds contain $CuO$ chains and $Cu-O-Cu$ angle
$\theta$ is near $90^{\circ}$ \cite{i1}. In this case the usual
antiferromagnetic super-exchange of two nearest-neighbor magnetic $Cu$ ions
is suppressed and the exchange integral $J_{12}$ is negative. On the other 
hand the next-nearest-neighbor interaction $J_{13}$ between $Cu$ ions
does not depend on $\theta$ and is antiferromagnetic. The estimation of
the ratio $J_{13}/J_{12}$ for $La_6Ca_8 Cu_{24}O_{41}$ and $Li_2CuO_2$
gives $\delta =0.11$ and $\delta =0.37$ respectively \cite{i1}.

Though the model (\ref{b19}) is not exactly solvable in this case, 
its properties for $\delta\ll 1$ can be studied. The consideration is based
on the classical approximation. In this approximation the ground state
spin structure is the spiral with the period $\sim\delta^{-1/2}$ and the
ground state energy $E=-2N\delta^2$. Using this approach the regular
expansion in powers of the small parameter $\delta$ can be developed
\cite{ko,zp}. The second-order quantum corrections coincides with the
classical energy. The calculation of higher orders of the perturbation theory
in $\delta$ leads to infrared-divergent integrals and it is necessary to sum
them in all orders to obtain the contributions proportional to $\delta^{5/2}$,
$\delta^3$ etc. In particular, the ground state energy calculated up to terms
$\sim\delta^{5/2}$ is
\[
E_0 = -4N\delta^2 + 4.14N\delta^{5/2}
\]

The excitation spectrum is gapless and has a sound like behaviour \cite{ko}. 
Corresponding sound velocity is $v=4\delta^{3/2}$.

The study of the dependence of the ground state energy $E(m)$ on magnetization
$m=S_z/N$ \cite{mag} at $\delta\ll 1$ shows that 
$\frac{\partial^2E}{\partial m^2}<0$ in a finite interval of values of $m$.
This implies the thermodynamic instability of the uniform state against the
phase separation. This instability arises due to existence of multimagnon
bound states. Energy of $n$-magnon bound state ($n\gg 1$) is
$\varepsilon_n=-nb$, $b=|\varepsilon_1|+\epsilon_b$, where $\varepsilon_1$ is
the energy of one-magnon state and $\epsilon_b$ is binding energy per
magnon. For the model (\ref{b19}) $\varepsilon_1=-8\delta^2$ and
$\epsilon_b\sim\delta^{5/2}$. As a result the function $E(m)$ has the form
\cite{mag}:
\begin{eqnarray}
E(m) &=& E_0 + 2\pi\delta^{3/2}m^2, \qquad m<m_c=\frac{2}{\pi}\sqrt{\delta}
\nonumber\\
E(m) &=& -\frac b2 (1-2m), \qquad m_c<m<\frac 12
\label{b20}
\end{eqnarray}
At $m>m_c$ the system is in a two-phase state consisting of the ferromagnetic
($m=1/2$) phase and the phase with $m=m_c$. According to (\ref{b20}) the
magnetization as a function of magnetic field $h$ has a jump from $m=m_c$ to
$m=1/2$ (metamagnetic transition) at $h=h_c=b$. These values $m_c$ and $h_c$
are close to those obtained in \cite{aligia} by extrapolation of finite
cluster calculations. We stress that the jump of the magnetization exists if
there are multimagnon bound states leading to the linear dependence of $E(m)$
for some region of $m$.

As for properties of the model far from the transition point 
$\delta =0$, now-days there is no clear understanding \cite{3,4,5}, but 
we believe that the spectrum remains gapless. For the case $\delta\gg 1$ this 
fact is predicted in \cite{12}. As for the jump in $m(h)$ it is unknown if it
remains at $\delta\simeq 1$. Numerical calculations \cite{aligia} show that,
at least, there is an abrupt increase of magnetization at some critical value
of magnetic field.

This perturbation theory can be generalized for the vicinity of 
any point on the transition line. It gives the similar properties of the
system in the region $J_{23}<0$, but it diverges at $J_{23}>0$.

At $J_{13}=0$ and $J_{23}>0$ the model (\ref{a1}) reduces to the
alternating Heisenberg chain studied in \cite{29}. The lowest excitation
is the triplet and there is the gap. At $J_{23}=0$ and $J_{13}>0$ the model
(\ref{a1}) reduces to the spin ladder with antiferromagnetic interactions
along legs and the ferromagnetic interactions on rungs. It is evident that
there is a gap at $J_{13}\ll 1$ (in this case the model is equivalent to the
spin $S=1$ Heisenberg chain). It was shown in \cite{30} that the gap
exists at $J_{13}\gg 1$.

The region $J_{23}>0,\;J_{13}<0$ was studied by different methods in 
\cite{epj,mag}. The exact diagonalization of finite chains shows the 
gap $\Delta$ in the excitation spectrum which is closed on the transition line.
The presence of singlet-triplet gap is also confirmed by calculations 
with use of variational wave function in MP form (\ref{d1}) when all matrix 
elements are considered as variational parameters. The MP 
variational function gives very good accuracy \cite{22,mag}, besides it 
gives exact ground state results for special points on the transition 
line and for the line $J_{13}=-\frac 12$ (dimer state).

\begin{figure}[tbp]
\unitlength1cm
\begin{picture}(8,5)
\centerline{\psfig{file=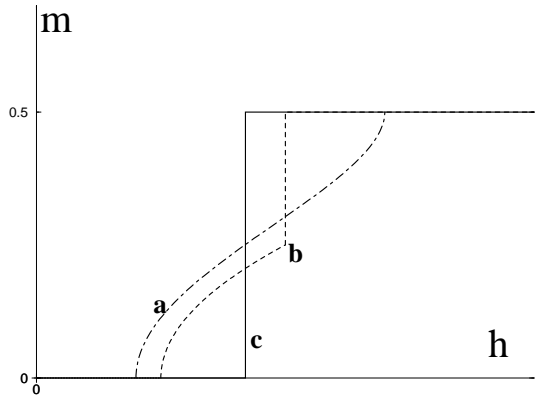,angle=-90,width=7cm}}
\end{picture}
\vspace{5mm}
\caption{ \label{mh} Dependence of magnetization $m$ on magnetic
field $h$ for the case:
(a) outside bound state region;
(b) in the bound state region, but outside the stripe 
$-\frac 12<J_{13}<-\frac 16$;
(c) inside the stripe $-\frac 12<J_{13}<-\frac 16$.}
\end{figure}

Another important property of the model with $J_{23}>0,\;J_{13}<0$ is the
existence of the multimagnon bound states for definite region of the values
$J_{23}$ and $J_{13}$. This region is shown on Fig.1. On the boundary of this
region the multimagnon bound states disappear. As it was noted before the
existence of the bound states leads to the linear regions on the dependence 
$E(m)$ and to the metamagnetic transition. The dependence $E(m)$ and $m(h)$
was found using MP variational function. The calculations show that out of the
bound state region the function $m(h)$ is typical for the gapped
antiferromagnet, i.e. it is monotonically increasing function (Fig.\ref{mh}). 
In other words,
$m(h)=0$ for $h<\Delta$ and $m(h)=1/2$ for $h>|\varepsilon_1|$. In the bound
state region the magnetization curve has the jump. It is interesting that
inside the stripe $-\frac 12<J_{13}<-\frac 16$ this jump takes 
place immediately from zero to the maximal value $m=\frac 12$ (Fig.\ref{mh})
at $h_c=|\varepsilon (0)|$ ($\varepsilon (0)$ is a ground state energy per
site). Such behaviour of $m(h)$ is explained by the fact that inside this
stripe $\Delta >|\varepsilon (0)|$  and the transition from the singlet state
to the ferromagnetic one comes to pass through states with intermediate values
of spin. In the bound state region outside the stripe the magnetization curve
has a form as shown on Fig.(\ref{mh}). In this case the critical field
$h_c=\Delta$. 

Thus, summarizing all above, we expect that the phase diagram of the model 
(\ref{a1}) has the form shown in Fig.(\ref{ph2}).

\section{Spin ladder model}

\begin{figure}[t]
\unitlength1cm
\begin{picture}(8,6)
\centerline{\psfig{file=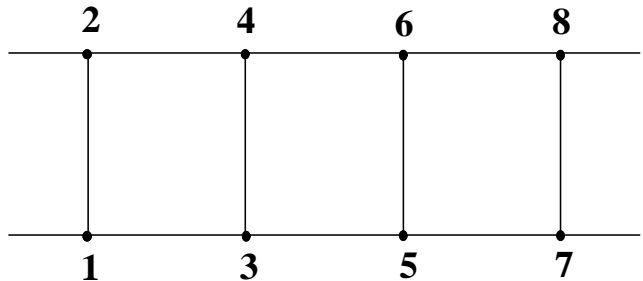,angle=-90,width=9cm}}
\end{picture}
\vspace{-20mm}
\caption{ \label{lad} The two-leg spin ladder.}
\end{figure}

Let us consider more general spin ladder model, for which `zigzag' model 
(\ref{a2}) is a particular case. So, we consider the cyclic ladder model 
containing $N=2M$ spins $s=\frac 12$ (Fig.\ref{lad}). The proposed form of 
wave function (\ref{a4}) can be generalized for a ladder model as follows:
\begin{equation}
\Psi_0 = P_0\Psi ,\qquad
\Psi = \langle 0|\,g_1\otimes g_2\otimes \ldots \otimes g_M \, |0 \rangle ,
\label{c1}
\end{equation}
where each $g_i$ corresponds to the $i$th rung of the ladder:
\begin{equation}
g_i = -b^+ (2x-b^+b) \left|\uparrow\uparrow \right \rangle _i
      + b \left|\downarrow\downarrow \right \rangle _i
      +(b^+b-x)\left|\uparrow\downarrow + \downarrow\uparrow\right \rangle _i
      +y \left|\uparrow\downarrow - \downarrow\uparrow\right \rangle _i 
\label{c2}
\end{equation}
where $x$ and $y$ are two parameters of the model. It is obvious that the
ladder wave function (\ref{c1})-(\ref{c2}) reduces to the original function 
(\ref{a4}) at $y=x$.

Now we will construct the Hamiltonian for which $\Psi_{\rm ladder}$ is the 
exact ground state wave function. This Hamiltonian describes two-leg 
$s=\frac 12$ ladder with periodic boundary conditions (Fig.\ref{lad}) 
and can be represented in a form
\begin{equation}
H=\sum\limits_{i=1}^{N}h_{i,i+1}, 
\label{c3}
\end{equation}
where $h_{i,i+1}$ describes the interaction between neighboring rungs. 
The spin space of two neighboring rungs consists of six multiplets: 
two singlet, three triplet and one quintet. At the same time, one can check 
that the product $g_i\otimes g_{i+1}$ contains only three of the six multiplets
of each pair of neighboring rungs: one singlet, one triplet and one quintet. 
The specific form of the singlet and triplet components present in the 
product $g_i\otimes g_{i+1}$ depends on parameters $x$ and $y$. 
The Hamiltonian 
$h_{i,i+1}$ can be written as the sum of the projectors onto the three missing
multiplets with arbitrary positive coefficients $\lambda_1,\,\lambda_2,\,
\lambda_3$: 
\begin{equation}
h_{i,i+1}=\sum\limits_{k=1}^{3}\lambda_{k}P_{k}^{i,i+1},  
\label{c4}
\end{equation}
where $P_k^{i,i+1}$ is the projector onto the missing multiplets in the
corresponding cell Hamiltonian. 

The wave function (\ref{c1}) is an exact wave function of the ground state
of the Hamiltonian $h_{i,i+1}$ with zero energy, because 
\begin{equation}
h_{i,i+1}|\Psi\rangle =0,\qquad i=1,...N-1  
\label{c5}
\end{equation}
and $\lambda_1,\,\lambda_2,\,\lambda_3$ are the excitation energies
of the corresponding multiplets.

So, $\Psi$ is the exact ground state wave function with zero energy for
the total Hamiltonian of an open ladder
\begin{equation}
H_{\rm open}=\sum_{i=1}^{M-1}h_{i,i+1}  
\label{c6}
\end{equation}
\begin{equation}
H_{\rm open}|\Psi\rangle =0  
\label{c7}
\end{equation}

Since the function $\Psi$ contains components with all possible values 
of total spin $S$ ($0\leq S\leq M$), then the ground state of open ladder 
is multiple degenerate. But for a cyclic ladder 
(\ref{c3}) only singlet and ferromagnetic components of $\Psi$ have zero
energy. Therefore, for a cyclic ladder $\Psi_0$ is 
a singlet ground state wave function degenerated with ferromagnetic state.
Besides, all the statements given for `zigzag' model (\ref{a2}) are valid 
for ladder model (\ref{c3}).

Since the specific form of the existing and missing multiplets in the wave
function (\ref{c1}) on each two nearest neighbor spin pairs depends on the
parameters $x$ and $y$, the projectors in (\ref{c4}) also
depend on $x$ and $y$. Each projector can be written in the form 
\begin{eqnarray}
P_{k}^{1,2} &=&J_{12}^{(k)}({\bf S}_{1}\cdot {\bf S}_{2}+{\bf S}_{3}\cdot 
{\bf S}_{4})+J_{13}^{(k)}({\bf S}_{1}\cdot {\bf S}_{3}+{\bf S}_{2}\cdot {\bf 
S}_{4})+J_{14}^{(k)}{\bf S}_{1}\cdot {\bf S}_{4}+J_{23}^{(k)}{\bf S}
_{2}\cdot {\bf S}_{3}  \nonumber \\
&+&J_{1}^{(k)}({\bf S}_{1}\cdot {\bf S}_{2})({\bf S}_{3}\cdot {\bf S}
_{4})+J_{2}^{(k)}({\bf S}_{1}\cdot {\bf S}_{3})({\bf S}_{2}\cdot {\bf S}
_{4})+J_{3}^{(k)}({\bf S}_{1}\cdot {\bf S}_{4})({\bf S}_{2}\cdot {\bf S}
_{3})+C^{(k)}  
\label{c8}
\end{eqnarray}
and this representation is unique for a fixed value of the parameters $x$ 
and $y$.

Substituting the above expressions for the projectors into Eq.~(\ref{c4}), 
we obtain the general form of the Hamiltonians $h_{i,i+1}$. Inasmuch as the
Hamiltonians $h_{i,i+1}$ have the same form for any $i$, it suffices
here to give the expression for~$h_{1,2}$:
\begin{eqnarray}
h_{1,2}=&&J_{12}(A_{12}+A_{34})+J_{13}(A_{13}+A_{24})+J_{14}A_{14}+J_{23}A_{23}
\nonumber \\
&+&J_{1}A_{12}A_{34}+J_{2}A_{13}A_{24}+J_{3}A_{14}A_{23}
\label{c9}
\end{eqnarray}
where
\[
A_{ij}={\bf S}_{i}\cdot {\bf S}_{j}-\frac{1}{4}
\]
and all exchange integrals depend on the model parameters and the spectrum
of excited states $J_{i}=J_{i}(x,y,\lambda_1,\lambda_2,\lambda_3)$ as follows:
\begin{eqnarray}
J_{12}&=&-\frac{\lambda_2}{2}+\frac{\lambda_3}{2} \frac{4y^2-1}{4y^2+1} ,
\qquad\qquad\quad
J_{23} = -\frac{\lambda_2}{2}-\frac{\lambda_3}{2} \frac{(2y-1)^2}{4y^2+1},
\nonumber \\
J_{13}&=&-\frac{\lambda_2}{2}-\frac{\lambda_3}{2} \frac{4y^2-1}{4y^2+1} ,
\qquad \qquad\quad 
J_{14} = -\frac{\lambda_2}{2}-\frac{\lambda_3}{2} \frac{(2y+1)^2}{4y^2+1},
\nonumber  \\ 
J_{1}&=&2J_{12} - \lambda_1 \frac{y^4-x^2(x+1)^2}{3y^4+x^2(x+1)^2} ,\qquad
J_{2} = 2J_{13} +2\lambda_1 \frac{y^4+y^2x(x+1)}{3y^4+x^2(x+1)^2} ,
\nonumber  \\ 
J_{3}&=&J_{14}+J_{23}+2\lambda_1 \frac{y^4-y^2x(x+1)}{3y^4+x^2(x+1)^2} 
\label{c10} 
\end{eqnarray}
(one should keep in mind that only positive $\lambda_i$ can be substituted 
to these expressions). The model (\ref{c10}) has evident symmetry: 
the change of sign
of $y$ is equivalent to renumbering of sites $1\leftrightarrow 2,\;
3\leftrightarrow 4\ldots$. Therefore, we will consider only the case $y>0$.

In general, the Hamiltonian $h_{i,i+1}$ contains all the terms presented in
(\ref{c8}), but we can simplify it 
by setting, for example, $J_2=J_3=0$ and solving equations (\ref{c10}) 
for $\lambda_1,\lambda_2,\lambda_3$. All $\lambda_k$ turn
out to be positive in this case for any $x$ and $y$ except two
lines $y=0$ and $x=-1/2$, where ground state is multiple degenerated. 
The Hamiltonian $h_{i,i+1}$ in this case takes the form
\begin{equation}
h_{1,2}=J_{12}(A_{12}+A_{34})+J_{13}(A_{13}+A_{24})+J_{14}A_{14}
       +J_{23}A_{23}+J_{1}A_{12}A_{34}
\label{c11}
\end{equation}
\begin{eqnarray*}
J_{12}&=&-x(x+1)+y^2-\frac 12 ,\qquad
J_{14} = x(x+1)-y^2-y   ,\\ 
J_{13}&=&-x(x+1)-y^2      ,\qquad\qquad
J_{23} = x(x+1)-y^2+y   ,\qquad 
J_{1}=\frac{[x(x+1)-y^2]^2}{y^2}-1
\end{eqnarray*}

The calculation of the norm of (\ref{c1}) and the singlet ground state 
correlation functions can be performed in the similar way to the 
corresponding calculations for the case $y=x=-1/4$.
Therefore, we give here the final result for spin correlation
functions at $N\rightarrow \infty $
\begin{eqnarray}
\left\langle {\bf S}_{n}\cdot{\bf S}_{n+2l}\right\rangle &=&
\frac 14 \cos \left( \frac{4\pi l}{N}\right), \nonumber \\
\left\langle {\bf S}_{2n-1}\cdot{\bf S}_{2n+2l}\right\rangle &=&
\frac 14 \cos \left( \frac{4\pi l}{N}-\triangle \varphi \right) 
\label{c13}
\end{eqnarray}
These equations mean that the spiral on each leg with pitch angle 
$\frac{4\pi }{N}$ is formed and the shift angle between spirals on the 
upper and the lower legs is $\triangle \varphi =\frac{8\pi y}{N}$.
At $y=0$, when spins on each rung form a local triplet, the shift angle 
vanishes and the spirals on both legs become coherent (we note that 
the shift angles in Eqs.(\ref{b18}) and (\ref{c13}) are defined in a 
different way).

Thus, there is just one full rotation of the spin over the 
length of the ladder, independent of the size of the system and for fixed 
$l\ll N$ at $N\rightarrow \infty $ two spins on the ladder are parallel.

We emphasize that the spin correlation function
$\langle{\bf S}_i\cdot{\bf S}_j\rangle$ does not depend on the choice of 
$\lambda_k$ for a fixed parameters $x,\,y$, because the ground state 
wave function of the three-parameter set of Hamiltonians (\ref{c10}) 
is the same.

\subsection{Special cases}

There are special values of the parameter $x$ for which Eqs.(\ref{c13}) 
are not valid. 
For cases of integer or half-integer $x=j$, which correspond to the 
special cases of the model (\ref{c10}), in Eq.(\ref{c2}) one can easily 
recognize the Maleev's boson representation of spin $S=j$ operators \cite{31}:
\begin{equation}
S^+=b^+(2j-b^+b),\qquad S^-=b, \qquad S^z=b^+b-j 
\label{d0}
\end{equation}
Generally, the wave functions (\ref{a4}) and (\ref{c1}) resemble MP form 
but with an infinite matrices represented by the Bose operators. 
However, in accordance to Maleev's representation in the special cases 
the infinite matrices formed by the Bose operators $b^{+}$ and $b$ can be 
broken off to the size $n=2j+1$ and wave function (\ref{c1}) is reduced to 
the usual MP form
\begin{equation}
\Psi_0=Tr\,\left( g_1\otimes g_2\otimes\ldots\otimes g_N\right) ,
\label{d1}
\end{equation}
where $g=-xT+yS$ is the $n\times n$ matrix describing states of spin pair 
on corresponding rung of the ladder. Singlet state matrix is
\begin{equation}
S=I\;\left| s\right\rangle   
\label{d2}
\end{equation}
where $I$ is unit matrix and $\left| s\right\rangle $ is the singlet
state. Triplet state matrix $T$\ is expressed by Clebsch-Gordan coefficients 
$C_{m_1,m_2}=\left\langle\left( 1,m_1\right)\left( j,m_2\right)\|
\left( j,m_1+m_2\right) \right\rangle$ as follows:
\begin{equation}
T=\frac{1}{C_{0,j}}\left( 
\begin{array}{ccccc}
C_{0,j}\left| 0\right\rangle  & C_{1,j-1}\left| 1\right\rangle  & 0 & 0 & 0
\\ 
C_{-1,j}\left| -1\right\rangle  & C_{0,j-1}\left| 0\right\rangle  & . & 0 & 0
\\ 
0 & . & . & . & 0 \\ 
0 & 0 & . & . & C_{1,-j}\left| 1\right\rangle  \\ 
0 & 0 & 0 & C_{-1,-j+1}\left| -1\right\rangle  & C_{0,-j}\left|
0\right\rangle 
\end{array}
\right) ,  
\label{d3}
\end{equation}
where $\left| \sigma \right\rangle $ is the triplet state with $S^{z}=\sigma $. 

Exact calculation of the correlators in the thermodynamic limit is performed
using standard transfer matrix technique and results in
\begin{eqnarray}
\left\langle {\bf S}_{1}{\bf S}_{2}\right\rangle  &=&
\frac{x(x+1)-3y^2}{4\omega}  ,
\nonumber \\
\left\langle {\bf S}_{i}{\bf S}_{i+2l}\right\rangle  &=&
\frac{x(x+1)(4y^2-1)}{4\omega^2}
\left( \frac{\omega -1}{\omega}\right)^{l-1}  ,
\nonumber \\
\left\langle {\bf S}_{2i+1}{\bf S}_{2i+2l+2}\right\rangle  &=&
-\frac{x(x+1)(2y-1)^2}{4\omega^2}
\left( \frac{\omega -1}{\omega}\right)^{l-1}  ,
\nonumber \\
\left\langle {\bf S}_{2i+2}{\bf S}_{2i+2l+1}\right\rangle  &=&
-\frac{x(x+1)(2y+1)^2}{4\omega^2}
\left( \frac{\omega -1}{\omega}\right)^{l-1}  ,
\label{d4}
\end{eqnarray}
where
\[
\omega=x(x+1)+y^2
\]

In the particular case of zero singlet weight, $y=0$, when spins on each
rung form a local triplet, correlation functions (\ref{d4}) coincide with
those obtained in \cite{27,28}.

According to Eqs.(\ref{d4}) the spin correlations have an exponential decay 
and the correlation length $r_{c}$ is
\begin{equation}
r_{c}=2\ln ^{-1}\left| \frac{\omega}{\omega -1}\right|   
\label{d5}
\end{equation}
In particular, for special points of `zigzag' model (\ref{a2}) with 
$J_{23}=\frac{2x}{2x+1}$ and $J_{13}=-x$ the correlation length
$r_c = -2 \ln ^{-1}\left( 1-\frac{1}{x(2x+1)} \right)$.

\begin{figure}[t]
\unitlength1cm
\begin{picture}(8,6)
\centerline{\psfig{file=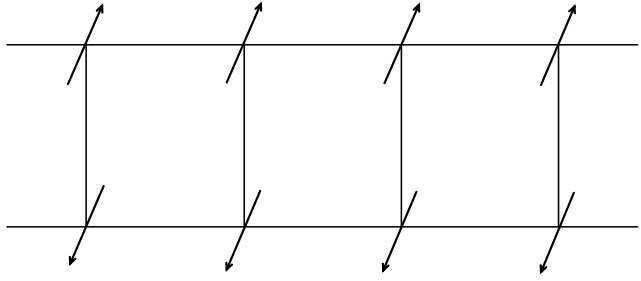,angle=-90,width=9cm}}
\end{picture}
\vspace{-20mm}
\caption{ \label{stripe} Stripe spin structure on the ladder model.}
\end{figure}

The correlation length $r_{c}$diverges when $x\rightarrow \infty$ or 
$y\rightarrow \infty$. In these cases the singlet ground state has collinear 
or stripe spin structure, i.e. spin-spin correlations are ferromagnetic along
legs and antiferromagnetic between them (Fig.\ref{stripe}), 
with a magnetic order $m$:
\[
\left \langle {\bf S}_{i}{\bf S}_{i+2l}  \right\rangle \simeq
-\left\langle {\bf S}_{i}{\bf S}_{i+1+2l}\right\rangle \simeq m^{2} ,
\qquad m\simeq\frac{xy}{x^2+y^2}
\]
When $y=x$ (the `zigzag' model) the magnetic order is equal to the 
classical value $m=1/2$.

We note that the wave function (\ref{c1}) show double-spiral ordering for
all values of $x$ and $y$ excluding the special lines $x=j$. The crossover 
between spiral and stripe states occurs in the exponentially small
(at $N\rightarrow \infty $) vicinity of the special lines.

\subsection{Spectrum of the model}

Generally, the excitation spectrum of model (\ref{c3},\ref{c11}) can not
be calculated exactly. It is clear that this spectrum is gapless because, for
example, the one-magnon energy is $\sim N^{-4}$ at $N\rightarrow \infty $.
Moreover, the lowest singlet excitation is gapless as well. For model 
(\ref{c3},\ref{c11}) lying on the special lines this fact can be established 
from the following consideration. As mentioned above, the crossover between 
spiral and stripe states occurs in the exponentially small vicinity of the 
special lines. It means that two wave functions $\Psi_0(j)$ and 
$\Psi_0(j+\delta)$ corresponding to points ($x=j,\;y$) and ($x=j+\delta,\;y$) 
respectively are almost orthogonal at $\delta\sim e^{-N}$. Hence, we can
consider a spiral wave function $\Psi_0(j+\delta)$ as a variational function 
of the excited singlet state at special point $x=j$. The energy of this 
excited state at special point $x=j$ is
\begin{eqnarray*}
E_{ex} &=& \left\langle \Psi_0(j+\delta )|\, 
                H(j)\, |\Psi_0(j+\delta )\right\rangle  \\
  &\sim &  \left\langle \Psi_0(j+\delta )|\,
            H(j+\delta )-\delta\frac{dH(j+\delta )}{dx}\,
                       |\Psi_0(j+\delta )\right\rangle
    \sim {\rm const}.\cdot\delta \sim e^{-N}
\end{eqnarray*}

Thus, on the special lines the stripe ground state of considered model is
asymptotically degenerated with an excited spiral singlet state at the 
thermodynamic limit. It is not clear if the degeneracy is exponentially 
large or not.
This consideration is valid for any integer or half-integer $x=j$,
but it is not valid for the parameters $x\neq j$, which are out of special 
lines. We performed \cite{epj} the numerical diagonalization of finite 
ladders for various parameters $x$ and $y$ and found that the exponential 
degeneracy possibly takes place for all parameters $x$ and $y$, 
but we can not confirm it strictly.
%

It is interesting to note that singlet wave function (\ref{c1})
can be also represented in a special recurrent form \cite{zp,prb}
\begin{equation} 
\Psi_0=P_0\Psi_M , 
\label{d7} 
\end{equation}
\begin{eqnarray}
\Psi_M &=&(s_1^+ +\nu_1 s_2^+ +\nu_2 s_3^+\ldots +\nu_2 s_{N}^+)
(s_3^+ +\nu_1 s_4^+\ldots +\nu_2 s_{N}^+)\ldots \nonumber \\
&&(s_{2n-1}^+ +\nu_1 s_{2n}^+\ldots +\nu_2 s_{N}^+)\ldots 
(s_{N-1}^+ +\nu_1 s_{N}^+) \mid\downarrow\downarrow\ldots\downarrow\rangle   
\label{d8}
\end{eqnarray}
where $s_i^+$ is the $s=\frac 12$ raising operator.
Eq.(\ref{d8}) contains $M$ operator multipliers and the vacuum
state $\mid\downarrow\downarrow\ldots\downarrow\rangle$ is the state with
all spins pointing down. The function $\Psi_M$ is the eigenfunction of 
$S_z$ with $S_z=0$ but it is not the eigenfunction of ${\bf S}^2$. 
$P_0$ is a projector onto the singlet state.
Two parameters $\nu_1$ and $\nu_2$ in wave function (\ref{d8}) is connected 
with parameters $x$ and $y$ in (\ref{c2}) as
\[
\nu_1=\frac{1+x-y}{1+x+y}, \qquad \nu_2=\frac{1}{1+x+y}
\]

The norm of the wave function (\ref{d7}) and expectation values can be also
calculated with use of recurrent technique developed in \cite{zp,prb}.
Certainly, it gives the same expressions (\ref{c13},\ref{d4}) 
for spin correlation function.

\subsection{Antiferromagnetic ladder model}

Now we should consider in particular the special case $x=1/2$ \cite{prb}.
In this case the product $g_i\otimes g_{i+1}$ contains only one singlet
and one triplet and does not contain quintet. Therefore, in this 
case the cell Hamiltonian can be written in the form \cite{jetp}
\begin{equation}
H = \sum_i h_{i,i+1} ,\qquad
h_{i,i+1}=\sum\limits_{k=1}^{4}\lambda_k P_k^{i,i+1},  
\label{e1}
\end{equation}
where $P_4^{i,i+1}$ - is a projector onto the quintet state. If all 
$\lambda_k>0$, the ferromagnetic state has positive energy $E=M\lambda_4$
and wave function $\Psi_0$ (\ref{c1}) is now non-degenerate singlet 
ground state wave function for the Hamiltonian (\ref{e1}), 
while for Hamiltonian (\ref{c3},\ref{c4}) $\Psi_0$ is also 
exact ground state but degenerate with ferromagnetic state. 
Using a freedom in choice of
$\lambda_k$ we can exclude all four-spin interactions in $h_{i,i+1}$
\begin{eqnarray}
h_{i,i+1} &=&J_{12}\left( {\bf S}_{2i-1}\cdot {\bf S}_{2i}
            +{\bf S}_{2i+1}\cdot {\bf S}_{2i+2} \right)
            +J_{13}\left( {\bf S}_{2i-1}\cdot {\bf S}_{2i+1}
            +{\bf S}_{2i}\cdot {\bf S}_{2i+2} \right) \nonumber  \\
          &&+J_{14} {\bf S}_{2i-1}\cdot {\bf S}_{2i+2}
            +J_{23} {\bf S}_{2i}\cdot {\bf S}_{2i+1} + C
\label{e2}
\end{eqnarray}
and all exchange integrals $J_{ij}$ depend on one model parameter $y<3/2$
(this inequality is necessary to satisfy $\lambda_k>0$)
\begin{eqnarray}
J_{12}&=& \frac 32 (4y^2-1)       ,\qquad\quad
J_{14} = -y(3+2y)(2y-1)^2,\nonumber \\
J_{13}&=& -2y^2(4y^2-1)   ,\qquad
J_{23} = y(3-2y)(2y+1)^2 ,\qquad 
C  = 9y^2 + \frac 34
\label{e3}
\end{eqnarray}

It follows from Eq.(\ref{d4}) that the ground state has ultrashort-range
correlations with $r_c\sim 1$. For example, $r_c(y=0)=2\log ^{-1}3$, 
which coincides with correlation length of AKLT model. 
But at $y=\frac 12$ all correlations are zero except 
$\left\langle {\bf S}_{2i}\cdot{\bf S}_{2i+1}\right\rangle =-\frac 34$.
It implies that at $y=\frac 12$ model (\ref{e1}-\ref{e3}) has a dimer 
ground state.

The value $(\omega -1)$ changes the sign at $y=\frac 12$ and as it
follows from Eqs.(\ref{d4}) the correlators show the
antiferromagnetic structure of the ground state at 
$\frac 12\leq y \leq\frac 32$
while at $0\leq y \leq\frac 12$ there are ferromagnetic correlations
inside pairs $(1,2), (3,4),\ldots $ and the antiferromagnetic correlations
between the pairs.


The Hamiltonian (\ref{e1}-\ref{e3}) of the cyclic ladder has a singlet-triplet
gap $\Delta$ for finite $N$. It is evident that for $y=\frac 12$ the gap 
exists for 
$N\rightarrow \infty $ and $\Delta (\frac 12)=4$. The existence of the
finite gap at the thermodynamic limit in the range $0<y<\frac 32$ follows from
the continuity of the function $\Delta (y)$. It is also clear that 
$\Delta (y)$ at $N\rightarrow \infty $ vanishes at the boundary points 
$y=0$ and $y=\frac 32$ when the ground state is degenerate and there are
low-lying spin wave excitations.

Unfortunately, the method of the exact calculation of $\Delta (y)$ in the
thermodynamic limit is unknown. For the
approximate calculation $\Delta (y)$ we use the trial function of the
triplet state in the form
\begin{equation}
\Psi_t = \sum_n s_n^{+} e^{ikn} \Psi_0 , 
\label{e5}
\end{equation}

The trial function $\Psi_t$ gives $\Delta (y)$ which at $N\rightarrow\infty$
has minima at $k=\pi$ and $k=0$ for $0<y<\frac 12$ and $\frac 12<y<\frac 32$, 
respectively
\begin{eqnarray}
\Delta (y)&=&\frac{32y^2(4y^2+1)}{4y^2+3} ,\qquad  0<y<\frac 12
\nonumber \\
\Delta (y)&=&\frac{128y^2}{(4y^2+1)(4y^2+3)} , \qquad 
\frac 12<y<\frac 32 
\label{e6}
\end{eqnarray}

\begin{figure}[t]
\unitlength1cm
\begin{picture}(12,6)
\centerline{\psfig{file=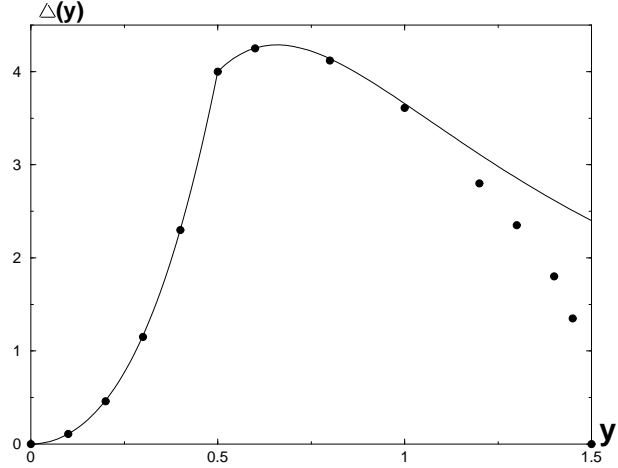,angle=-90,width=8cm}}
\end{picture}
\vspace{5mm}
\caption[]{Singlet-triplet gap of the model (\ref{e3}) as a function of the
parameter $y$. The circles denote the results of the extrapolation of an
exact finite-chain calculations. The solid line represents the dependence
$\Delta (y)$ given by Eq.(\ref{e6}).
\label{gap} }
\end{figure}

The dependence of $\Delta (y)$ given by Eq.(\ref{e6}) is shown on 
Fig.\ref{gap}
together with the results of extrapolations of exact finite-chain
calculations. Both dependences agree very well for $y\leq\frac 12$. 
However, $\Delta (y)$ given by Eq.(\ref{e6}) is not 
zero at $y=\frac 32$, while numerical calculations fit the dependence 
$\Delta (y)\sim\sqrt{\frac 32-y}$ at $y\rightarrow\frac 32$.

We note that the trial function of the type (\ref{e5}) gives the value $0.7407$
for the singlet-triplet gap in the AKLT model. This estimate is close to
the value $0.7143$ obtained by another approach in \cite{19}.


For other special cases one can also construct Hamiltonians for which 
$\Psi_0$ is non-degenerate singlet ground state wave function. 
But in these cases one have to introduce more distant interactions. 
%
%
%

\section{Valence-Bond-State models}

\subsection{One-dimensional model}

We studied previously a one-parameter ladder model (\ref{e1}-\ref{e3}) 
with non-degenerate singlet ground state. The exact ground state wave 
function of the cyclic ladder was written in the MP form (\ref{d1}).
Now we write the wave function $\Psi_0$ in a form more suitable for
subsequent generalization to other types of lattices \cite{jetp}.

We consider a ladder of $N=2M$ spins~1/2.  The wave function of this
system is described by the $N$th-rank spinor
\begin{eqnarray}
\Psi = \Psi ^{\lambda \mu \nu \ldots \tau },
\label{g1}
\end{eqnarray}
where the indices $\lambda ,\, \mu ,\, \nu ,\ldots,\, \tau = 1,\,2$
correspond to different projections of the spin~1/2.

We partition the system into pairs of spins located on rungs of the ladder.
The wave function can then be written as the product of $M$ second-rank spinors
\begin{eqnarray}
\Psi = \Psi^{\lambda\mu}(1)\Psi^{\nu\rho}(2)\ldots\Psi^{\sigma\tau}(M).
\label{g2}
\end{eqnarray}
We now form a scalar from Eq.~(\ref{g2}), simplifying the latter with
respect to index pairs:
\begin{eqnarray}
\Psi_{s} = \Psi^{\lambda}{}_{\nu}(1)\Psi^{\nu}{}_{\kappa}(2)\ldots
\Psi^{\sigma}{}_{\lambda}(M).
\label{g3}
\end{eqnarray}
Here subscripts correspond to the covariant components of the spinor, which
are related to the contravariant components (superscripts) through the
metric spinor
\begin{eqnarray}
g_{\lambda \mu } = g^{\lambda \mu } = \left (
\begin{array}{cc}
 0 & 1 \\
-1 & 0 \\
\end{array}
\right ). \label{g4} \\
\Psi _{\lambda } = g_{\lambda \mu }\Psi ^{\mu }, \qquad \Psi ^{\lambda } =
g^{\mu \lambda }\Psi _{\mu } .\nonumber
\end{eqnarray}

The scalar function (\ref{g3}) can thus be written in the form
\begin{eqnarray}
\Psi_s = \Psi^{\lambda\mu}(1)g_{\mu\nu}\Psi^{\nu\rho}(2)g_{\rho
\kappa}\ldots\Psi^{\sigma\tau}(M)g_{\tau\lambda}.
\label{g5}
\end{eqnarray}
The scalar function $\Psi_s$ obviously describes to the singlet state.

The second-rank spinor describing the pair of spins~1/2 can be written in
the form
\begin{eqnarray}
\Psi^{\lambda\mu} = c_t\Psi_t^{\lambda\mu}+c_s\Psi_s^{\lambda\mu},
\label{g6}
\end{eqnarray}
where $\Psi_t^{\lambda\mu}$ and $\Psi_s^{\lambda\mu}$ are
symmetric and antisymmetric second-rank spinors, respectively, and $c_t$
and $c_s$ are arbitrary constants.  We know that the symmetric
second-rank spinor describes a system with spin $S=1$, so that the pair of
spins~1/2 in this case forms a triplet.  If $\Psi^{\lambda\mu}$ is an
antisymmetric second-rank spinor reducible to a scalar multiplied by
$g_{\lambda\mu}$, the spin pair exists in the singlet state.
Consequently, the ratio of the constants $c_t$ and $c_s$ determines the
relative weights of the triplet and singlet components on the pair of spins
$s = 1/2$ and is a parameter of the model.  In particular, for $c_s=0$
the wave function (\ref{g6}) contains only a triplet component, and for
$c_t=0$ it contains only a singlet component.

%
We note that the wave function (\ref{g3}) has MP form (\ref{d1})
with the matrices $g_i$ representing a mixed second-rank tensor:
\begin{eqnarray}
g_i = \Psi^{\lambda}{}_{\nu}(i) = c_t\left (
\begin{array}{cc}
{\displaystyle \frac 12 
\left|\uparrow\downarrow + \downarrow\uparrow\right \rangle_i }
& \left|\downarrow\downarrow\right \rangle_i \\
- \left|  \uparrow  \uparrow\right \rangle_i
& {\displaystyle -\frac 12
\left|\uparrow\downarrow + \downarrow\uparrow\right \rangle_i } \\
\end{array} \right ) 
- \frac 12 c_s(
\left|\uparrow\downarrow + \downarrow\uparrow\right \rangle_i \,I,
\label{g7}
\end{eqnarray}
where $I$ is the unit matrix.

We now choose a Hamiltonian $H$ for which the wave function (\ref{g5})
is an exact ground-state wave function.  To do so, we consider the part of
the system (cell) consisting of two nearest neighbor spin pairs.  In the
wave function (\ref{g5}) the factor corresponding to the two spin pairs is
a second-rank spinor:
\begin{eqnarray}
\Psi^{\lambda\mu}(i)g_{\mu\nu}\Psi^{\nu\rho}(i+1).
\label{g8}
\end{eqnarray}
In the general case, therefore, only two of the six multiplets forming two
pairs of spin~1/2 --- one singlet and one triplet --- are present in the
function (\ref{g8}).  Inasmuch as four spins~1/2 form two singlets
and three triplets, the specific form of the singlet and triplet components
present in the wave function (\ref{g8}) depends on the ratio $c_s/c_t$. 
The cell Hamiltonian acting in the spin space of nearest neighbor 
spin pairs can be written as the sum of the projectors onto the four missing 
multiplets with arbitrary positive coefficients $\lambda_k$ (\ref{e1}). 
As mentioned above, the general form of the Hamiltonian (\ref{e1}) can be 
reduced to a more simple form (\ref{e2}) with $c_s/c_t=2y$.

Thus, the singlet ground state wave function of the model (\ref{e1}) can be 
also written in a spinor form (\ref{g5}).


\subsection{Two-dimensional model}

\begin{figure}[t]
\unitlength1cm
\begin{picture}(11,6)
\centerline{\psfig{file=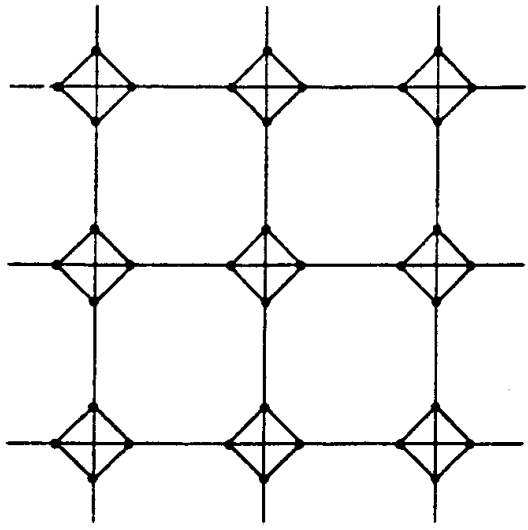,angle=0,width=8cm}}
\end{picture}
\caption[]{Two-dimensional lattice on which the spin model is
defined.\label{jetp1}}
\end{figure}

Now we consider an $N\times N$-site square lattice with cyclic boundary
conditions.  We replace each site of the lattice by a square
(Fig.\ref{jetp1}) with spins $s = 1/2$ at its corners, making the total
number of spins equal to $4N^2$.  To avoid misunderstanding, however,
from now on we continue to refer to these squares as sites.  The wave
function of the system is described by the product of fourth-rank spinors
\begin{eqnarray}
\Psi = \prod_{\bf n} \Psi^{\lambda_{\bf n}\mu_{\bf n}\nu_{\bf n}\rho_{\bf n}}
({\bf n}).
\label{g9}
\end{eqnarray}
By analogy with (\ref{g5}), from Eq.~(\ref{g9}) we form the scalar
\begin{eqnarray}
\Psi = \prod \limits _{{\bf n}}\Psi ^{\lambda _{{\bf n}}\mu _{{\bf n}}\nu
_{{\bf n}}\rho _{{\bf n}}}({\bf n})g_{\nu _{{\bf n}} \lambda _{{\bf n} +
{\bf a}}}g_{\rho _{{\bf n}} \mu _{{\bf n} + {\bf b}}}.
\label{g10}
\end{eqnarray}
where {\bf a} and {\bf b} are unit vectors in the {\it x} and {\it y}
directions.

\begin{figure}[t]
\unitlength1cm
\begin{picture}(11,6)
\centerline{\psfig{file=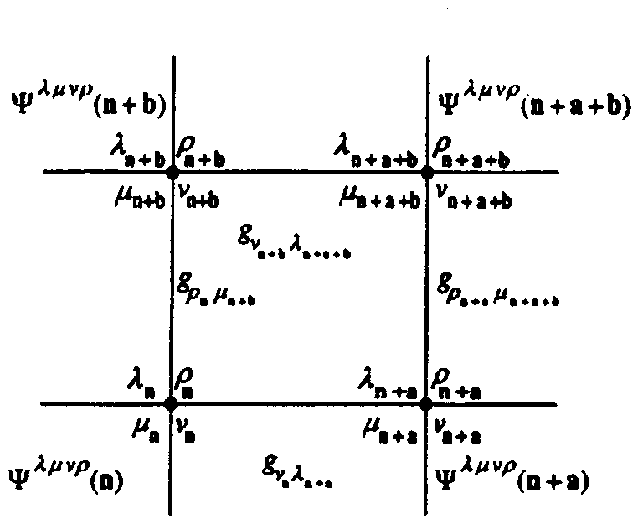,angle=0,width=8cm}}
\end{picture}
\caption[]{Graphical correspondence of the model wave function.  The
indices of the site spinors depend on the site index (not shown in the
figure).\label{jetp2}}
\end{figure}

The singlet wave function (\ref{g10}) is conveniently identified
graphically with a square lattice, each site corresponding to a fourth-rank
spinor $\Psi ^{\lambda \mu \nu \rho }$ (whose form is identical for all
sites), and each segment linking sites corresponds to a metric spinor
$g_{\lambda \mu }$ (Fig.\ref{jetp2}).

To completely define the wave function (\ref{g10}), it is necessary to know
the form of the site spinor $\Psi ^{\lambda \mu \nu \rho }$. 
The specific form of the fourth-rank spinor $\Psi ^{\lambda
\mu \nu \rho }$ [and, hence, the wave function (\ref{g10})] describing the
system of four spins $s = 1/2$ is governed by $14$ quantities
\cite{jetp}, which are parameters of the model.

We now choose a Hamiltonian {\it H} for which the wave function (\ref{g10})
is an exact ground state wave function.  As in the one-dimen\-sional case,
we seek the required Hamiltonian in the form of a sum of cell Hamiltonians
acting in the space of two nearest neighbor spin quartets:
\begin{eqnarray}
H = \sum \limits _{{\bf n}}H_{{\bf n},{\bf n} + {\bf a}} +
    \sum \limits _{{\bf n}}H_{{\bf n},{\bf n} + {\bf b}}.
\label{g14}
\end{eqnarray}

\begin{figure}[t]
\unitlength1cm
\begin{picture}(11,6)
\centerline{\psfig{file=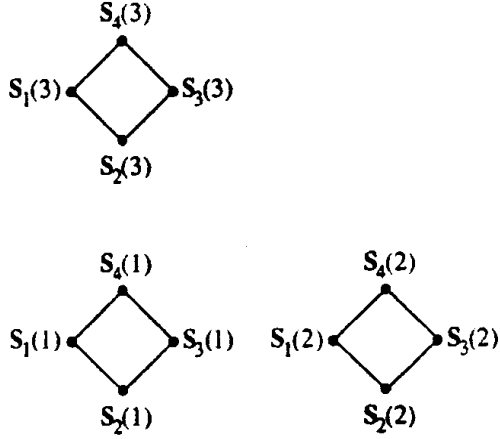,angle=0,width=8cm}}
\end{picture}
\caption[]{Lattice sites associated with interactions $H_{1,2}$
and~$H_{1,3}$.\label{jetp3}}
\end{figure}

The first term in Eq.~(\ref{g14}) is the sum of the cell Hamiltonians in
the horizontal direction, and the second term is the same for the vertical.
The cell Hamiltonians along each direction have the same form, but the
``horizontal'' and ``vertical'' Hamiltonians differ in general.  In the
following discussion, therefore, we consider only the Hamiltonians $H_{1,2}$
and $H_{1,3}$ (Fig.\ref{jetp3}), which describe interactions of ``sites'' in
the {\it x} and {\it y} directions, respectively.

For the wave function (\ref{g10}) to be an exact eigenfunction of the
Hamiltonian {\it H}, it is sufficient that the sixth-rank spinors
\begin{eqnarray}
&{}&\Psi ^{\lambda _{1}\mu _{1}\nu _{1}\rho _{1}}(1)
\Psi ^{\lambda _{2}\mu _{2}\nu _{2}\rho _{2}}(2)
g_{\nu _{1}\lambda _{2}},
\nonumber \\
&{}&\Psi ^{\lambda _{1}\mu _{1}\nu _{1}\rho _{1}}(1)
\Psi ^{\lambda _{3}\mu _{3}\nu _{3}\rho _{3}}(3)
g_{\rho _{1}\mu _{3}},
\label{g15}
\end{eqnarray}
be eigenfunctions of the corresponding cell Hamiltonians $H_{1,2}$
and~$H_{1,3}$.

In general, when the site spinor $\Psi ^{\lambda \mu \nu \rho }$ is not
symmetric with respect to any indices, the possible states of two quartets
of spins $s = 1/2$ consist of 70 multiplets.  A wave function represented
by a sixth-rank spinor contains only 20 of them.  Accordingly, the cell
Hamiltonians $H_{1,2}$ and $H_{1,3}$ can be represented by the sum of
projectors onto the 50 missing multiplets:
\begin{eqnarray}
H_{1,2} = \sum \limits _{k = 1}^{50}\lambda _{k}P_{k}^{1,2}, \qquad
H_{1,3} = \sum \limits _{k = 1}^{50}\mu _{k}P_{k}^{1,3},
\label{g16}
\end{eqnarray}
where the positive constants $\lambda _{k}$ and $\mu _{k}$ are the
excitation energies of $H_{1,2}$ and $H_{1,3}$, and the specific form of
the projectors depends on 14 model parameters.

Inasmuch as
\begin{eqnarray}
H_{{\bf n},{\bf n} + {\bf a}}|\Psi _{s}\rangle = 0, \quad
H_{{\bf n},{\bf n} + {\bf b}}|\Psi _{s}\rangle = 0,
\label{g17}
\end{eqnarray}
for the total Hamiltonian (\ref{g14}) we have the expression
\begin{eqnarray}
H|\Psi _{s}\rangle = 0.
\label{g18}
\end{eqnarray}

Consequently, $\Psi _{s}$ is the ground-state wave function of the total
Hamiltonian {\it H}, because it is a sum of nonnegative definite cell
Hamiltonians.  Also, it can be rigorously proved \cite{jetp} that
the ground state of {\it H} is nondegenerate.

As mentioned above, the specific form of the projectors depends on 14 model
parameters, and in general the cell Hamiltonians (\ref{g16}), expressed in
terms of scalar products of the type ${\bf s}_i\cdot{\bf s}_j$, 
$({\bf s}_i\cdot{\bf s}_j)({\bf s}_k\cdot{\bf s}_l)$, etc., have an
extremely cumbersome form.  We therefore consider a few special cases.

When the site spinor $\Psi^{\lambda\mu\nu\rho}$ is a symmetric
fourth-rank spinor $Q^{\lambda\mu\nu\rho}$ (corresponding to the
two-dimensional AKLT model\cite{40}), only the quintet component out of 
the six multiplets on each spin quartet is present in the wave function
(\ref{g10}).  The sixth-rank spinors (\ref{g15}) are
symmetric with respect to two triplets of indices and, hence, contain four
multiplets with $S = 0,\, 1,\, 2,\, 3$ formed from two quintets.
Consequently, the cell Hamiltonian ($H_{1,2}$ and $H_{1,3}$ coincide in
this case) has the form
\begin{eqnarray}
H_{1,2} = \sum \limits _{k = 1}^{66}\lambda _{k}P_{k}^{1,2}.
\label{g19}
\end{eqnarray}
If we set $\lambda _{k} = 1$ ($k = 1,\, 66$), we can write Eq.~(\ref{g19})
in the form
\begin{eqnarray}
H_{1,2} = P_{4}({\bf S}_{1} + {\bf S}_{2}) + \big [1 - P_{2}({\bf
S}_{1})P_{2}({\bf S}_{2})\big ],
\label{g20}
\end{eqnarray}
where ${\bf S}_{i}$ is the total spin of the quartet of spins $s = 1/2$ on
the {\it i}th site, ${\bf S}_{i} = {\bf s}_{1}(i) + {\bf s}_{2}(i) + {\bf
s}_{3}(i) + {\bf s}_{4}(i)$, and $P_{l}({\bf S})$ is the projector onto
the state with spin $S = l$.

If the four spins $s = 1/2$ at each site are replaced by a single spin $S =
2$ and if the wave function (\ref{g10}) is treated as a wave function
describing a system of $N^2$ spins $S = 2$, the second term in the
Hamiltonian (\ref{g20}) vanishes, and we arrive at the Hamiltonian of the
two-dimen\-sional AKLT model:
\begin{eqnarray}
H_{1,2} = P_{4}({\bf S}_{1} + {\bf S}_{2}) = \frac{1}{28}{\bf S}_{1}\cdot
{\bf S}_{2} + \frac{1}{40}({\bf S}_{1}\cdot {\bf S}_{2})^{2} +
\frac{1}{180}({\bf S}_{1}\cdot {\bf S}_{2})^{3} + \frac{1}{2520}({\bf
S}_{1}\cdot {\bf S}_{2})^{4}.
\label{g21}
\end{eqnarray}

Another interesting special case is encountered when the system decomposes
into independent one-dimen\-sional chains.  This happens if the site spinor
$\Psi ^{\lambda \mu \nu \rho }$ reduces to a product of two second-rank
spinors, each describing two spins~1/2.  For example,
\begin{eqnarray}
\Psi ^{\lambda \mu \nu \rho }(s_{1},\, s_{2},\, s_{3},\, s_{4}) = \varphi
^{\lambda \nu }(s_{1},\, s_{3})\varphi ^{\mu \rho }(s_{2},\, s_{4}).
\label{g22}
\end{eqnarray}
In this case the Hamiltonians $H_{1,2}$ and $H_{1,3}$ contain interactions
of four rather than eight spins~1/2 and have the form~(\ref{c9}).

\begin{figure}[t]
\unitlength1cm
\begin{picture}(11,6)
\centerline{\psfig{file=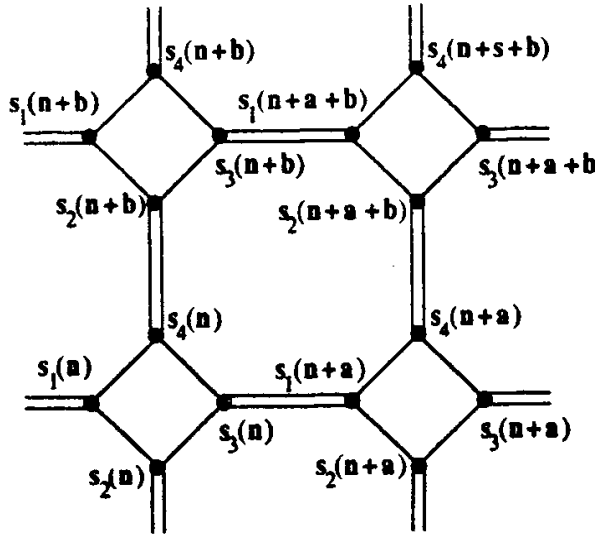,angle=0,width=8cm}}
\end{picture}
\caption[]{Pattern of independent singlet pairs (double lines).\label{jetp4}}
\end{figure}

The simplest case is when the site spinor $\Psi ^{\lambda \mu \nu \rho }$
is a product of four first-rank spinors:
\begin{eqnarray}
\Psi ^{\lambda \mu \nu \rho }(s_{1},\, s_{2},\, s_{3},\, s_{4}) = \varphi
^{\lambda }(s_{1})\varphi ^{\mu }(s_{2})\varphi ^{\nu }(s_{3})\varphi
^{\rho }(s_{4}).
\label{g23}
\end{eqnarray}
Now the system decomposes into independent singlet pairs (Fig.\ref{jetp4}),
and the total Hamiltonian of the system has the form
\begin{eqnarray}
H = \sum \limits _{i,j}\left ({\bf s}_{i}\cdot {\bf s}_{j} +
\frac{3}{4}\right ),
\label{g24}
\end{eqnarray}
where ${\bf s}_{i}$ and ${\bf s}_{j}$ are the spins forming the singlet
pairs.

\subsection{Spin correlation functions in the ground state}

We now look at the problem of calculating the norm and the correlation
function of the model described by the wave function (\ref{g10}).  The
expression for the norm of the wave function $G = \langle \Psi_s|\Psi
_{s}\rangle $ has the form
\begin{eqnarray}
G &=& \prod \limits _{{\bf n}}\left \langle
\Psi ^{\lambda '_{{\bf n}}\mu '_{{\bf n}}\nu '_{{\bf n}}\rho ' _{{\bf
n}}}({\bf n})\Big |
\Psi ^{\lambda _{{\bf n}}\mu _{{\bf n}}\nu _{{\bf n}}\rho
_{{\bf n}}}({\bf n})\right \rangle
g_{\nu _{{\bf n}}\lambda _{{\bf n} + {\bf a}}}
g_{\rho _{{\bf n}}\mu _{{\bf n} + {\bf b}}}
g_{\nu '_{{\bf n}}\lambda '_{{\bf n} + {\bf a}}}
g_{\rho '_{{\bf n}}\mu '_{{\bf n} + {\bf b}}} \nonumber \\ [3mm]
&=& \prod \limits _{{\bf n}}
R_{\lambda _{{\bf n}}\mu _{{\bf n}}\lambda _{{\bf n} + {\bf a}}\mu _{{\bf
n} + {\bf b}}}
^{\lambda '_{{\bf n}}\mu '_{{\bf n}}\lambda '_{{\bf n} + {\bf a}}\mu
'_{{\bf n} + {\bf b}}} =
\prod \limits _{{\bf n}}
R_{\alpha _{{\bf n}}\beta _{{\bf n}}\alpha _{{\bf n} + {\bf a}}\beta _{{\bf
n} + {\bf b}}},
\qquad \alpha _{i},\, \beta _{i} = \{1,\, 2,\, 3,\, 4\},
\label{h1}
\end{eqnarray}
where $R_{\alpha _{}\beta _{}\alpha _{}\beta _{}}$ is a $4\times 4\times
4\times 4$ matrix.

According to the selection rules for the projection of the total spin
$S^{z}$, only 70 of the 256 elements in the expression $\Bigl \langle \Psi
^{\lambda '_{}\mu '_{}\nu '_{}\rho _{}}({\bf n})\Big |\Psi ^{\lambda _{}\mu
_{}\nu _{}\rho _{}}({\bf n})\Big \rangle $ are non-vanishing.  Consequently,
the matrix {\it R} also contains at most 70 elements.  If we regard the
elements of {\it R} as Boltzmann vertex weights, the problem of calculating
the norm reduces to the classical 70-vertex model.
Since the exact solution for the 70-vertex model is unknown, numerical
methods must be used to calculate the norm and the expectation values.

To calculate the above-indicated expected values, we carry out Monte Carlo
calculations on $20\times 20$-site lattices.  As mentioned, the
ground-state wave function of the model depends on 14 parameters and, of
course, cannot possibly be analyzed completely.  We confine the numerical
calculations to the case in which the spinor $\Psi ^{\lambda \mu \nu \rho
}$ depends on one parameter~$\alpha $:
\begin{eqnarray}
\Psi ^{\lambda \mu \nu \rho } = \cos \alpha \cdot Q^{\lambda \mu \nu \rho }
+ \sin \alpha \cdot \Big (A^{\lambda \mu \nu \rho } - Q^{\lambda \mu \nu
\rho }\Big ),
\label{h2}
\end{eqnarray}
where $\alpha \in \big [-\pi /2;\,\, \pi /2\big ]$, the spinor $Q^{\lambda
\mu \nu \rho }$ is symmetric with respect to all indices, and
\begin{eqnarray}
A^{\lambda \mu \nu \rho } =
\varphi ^{\lambda }(s_{1})\varphi ^{\mu }(s_{2})
\varphi ^{\nu }(s_{3})\varphi ^{\rho }(s_{4}).
\label{h3}
\end{eqnarray}

In this case we have a one-parameter model with two well-known limiting
cases.  One corresponds to $\alpha  = \pi /4$, for which $\Psi ^{\lambda
\mu \nu \rho } = A^{\lambda \mu \nu \rho }$, and the system decomposes into
independent singlet pairs (Fig.~\ref{jetp4}); the other limiting case
corresponds to $\alpha = 0$ (our model reduces to the two-dimen\-sional
AKLT model in this case, the spins at each site forming a quintet).

In the given model there are four spins $s = 1/2$ at each site, and the
enumeration of each spin is determined by the order number of the lattice
site to which it belongs and by its own number at this site.  The spin
correlation function therefore has the form
\begin{eqnarray}
f_{ij}({\bf r}) = \big \langle {\bf s}_{i}({\bf n})\cdot {\bf s}_{j}({\bf
n} + {\bf r})\big \rangle .
\label{h4}
\end{eqnarray}

In determining the spin structure of the ground state, however, it is more
practical to consider the more straightforward quantity~$F({\bf r})$:
\begin{eqnarray}
F({\bf r}) = \sum \limits _{i,j = 1}^{4}\big \langle {\bf s}_{i}({\bf
n})\cdot {\bf s}_{j}({\bf n} + {\bf r})\big \rangle =
\big \langle {\bf S}({\bf n})\cdot {\bf S}({\bf n} + {\bf r})\big \rangle.
\label{h5}
\end{eqnarray}

\begin{figure}[t]
\unitlength1cm
\begin{picture}(11,6)
\centerline{\psfig{file=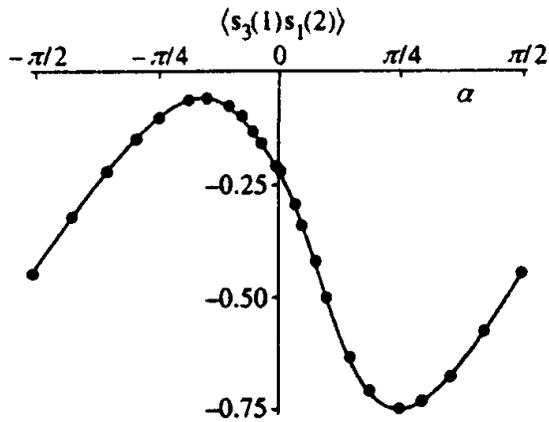,angle=0,width=8cm}}
\end{picture}
\caption[]{Dependence of the spin correlation function $\langle {\bf
s}_{3}(1){\bf s}_{1}(2)\rangle $ on the parameter~$\alpha $.\label{jetp5}}
\end{figure}

The function $F({\bf r})$ is left unchanged by a change of sign of $\alpha$.
We note, however, that only the total
correlation function, and not $f_{ij}({\bf r})$, possesses symmetry under a
change of sign of $\alpha $.  This assertion is evident, for example, in
Fig.~\ref{jetp5}, which shows the dependence of $f_{31}({\bf a})$ on $\alpha $
as an illustration.

\begin{figure}[t]
\unitlength1cm
\begin{picture}(11,6)
\centerline{\psfig{file=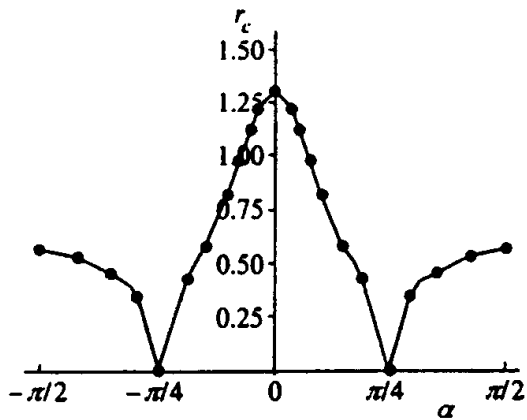,angle=0,width=8cm}}
\end{picture}
\caption[]{Dependence of the correlation length on the parameter~$\alpha$.
\label{jetp7}}
\end{figure}

The correlation
function decays exponentially as {\bf r} increases, differing from the
one-dimen\-sional model in that the pre-exponential factor also depends on
{\bf r}.  Figure~\ref{jetp7} shows the dependence of the correlation length
$r_{c}$ on the parameter $\alpha $.  The correlation length is a maximum at
the point $\alpha = 0$ (two-dimen\-sional AKLT model), decreases as $\alpha
$ increases, and at $\alpha = \pi /4$, when the system decomposes into
independent singlet pairs (Fig.~\ref{jetp4}), it is equal to zero.  With a
further increase in $\alpha $ the correlation length increases and attains
a second maximum at $\alpha = \pi /2$.  Like the correlation function
$F({\bf r})$, the function $r_{c}(\alpha )$ is symmetric with respect to
$\alpha $.  It is evident from Fig.~\ref{jetp7} that the parameter $\alpha $
has two ranges corresponding to states with different symmetries.  In the
range $|\alpha | < \pi /4$ the correlation function $F({\bf r})$ exhibits
antiferromagnetic behavior:
\begin{eqnarray}
F({\bf r}) \propto (-1)^{r_{x} + r_{y}}e^{-|{\bf r}|/r_{c}},
\label{h8}
\end{eqnarray}
whereas the spins at one site are coupled ferromagnetically, $\langle {\bf
s}_{i}({\bf n})\cdot {\bf s}_{j}({\bf n})\rangle > 0$.  On the other hand,
in the range $\pi /4 < |\alpha | < \pi /2$ the correlation function $F({\bf
r})$ is always negative:
\begin{eqnarray}
F({\bf r}) \propto -e^{-|{\bf r}|/r_{c}}
\label{h9}
\end{eqnarray}
and all the correlation functions at one site are also negative.
%

These ranges have two end points in common, $\alpha  = \pm \pi /4$, where
$r_{c} = 0$.  Whereas $\alpha = \pi /4$ corresponds to the trivial
partition of the system into independent singlet pairs, the case $\alpha
= -\pi /4$ is more interesting. In this case one can calculate all spin
correlations exactly \cite{jetp}. The correlations of spins located on 
neighboring `sites' of lattice at $\alpha =-\pi /4$ are antiferromagnetic, 
while all other correlators are zero.

The Hamiltonian for model (\ref{h2}) has a very combersome form and for
the cases $\alpha = -\pi /4$ and $\alpha = \pi /2$ was given in \cite{jetp}.

Our results suggest that the spin correlation functions decay exponentially
with a correlation length $\sim 1$ for an arbitrary parameter $\alpha $.
We also assume that the decay of the correlation function is of the
exponential type for the 14-parameter model as well, i.e., for any choice
of site spinor $\Psi ^{\lambda \mu \nu \rho }$.  This assumption is
supported in special cases: 
1)~the partition of the system into one-dimen\-sional chains with exactly 
known exponentially decaying correlation functions; 
2)~the two-dimen\-sional AKLT model, for which the
exponential character of the decay of the correlation function has been
rigorously proved \cite{41}.  
Further evidence of the stated assumption lies in the numerical results 
obtained for various values of the parameter in the one-parameter model.

\subsection{Generalization of the model to other types of lattices}

The wave function (\ref{g5}),\,\,(\ref{g10}) can be generalized to any type
of lattice.  The general principle of wave function construction for a
system of spins~1/2 entails the following:

1) Each bond on a given lattice has associated with it two indices running
through the values 1 and 2, one at each end of the bond.

2) Each bond has associated with it a metric spinor $g_{\lambda \mu }$ with
the indices of the ends of this bond.

3) Each site of the lattice (a site being interpreted here, of course, in
the same sense as in Sec.IVB) with {\it m} outgoing bonds has
associated with it an {\it m}th-rank spinor with the indices of the bonds
adjacent to the site.

4) The wave function is the product of all spinors at sites of the lattice
and all metric spinors.

It is obvious that each index in the formulated wave function is
encountered twice, so that the wave function is scalar and, hence, singlet.

The wave function so constructed describes a system in which each lattice
site contains as many spins $s = 1/2$ as the number of bonds emanating from
it.

To completely define the wave function, it is necessary to determine the
specific form of all site spinors.  The coefficients that determine their
form are then parameters of the model.

The Hamiltonian of such a model is the sum of the cell Hamiltonians acting
in the spin space of the subsystem formed by the spins at two mutually
coupled sites:
\begin{eqnarray}
H = \sum \limits _{\langle ij\rangle }H_{ij}.
\label{h17}
\end{eqnarray}

Each cell Hamiltonian is the sum of the projectors with arbitrary positive
coefficients onto all multiplets possible in the corresponding two-site
subsystem except those present in the constructed wave function:
\begin{eqnarray}
H_{i,j} = \sum \limits _{k}\lambda _{k}P_{k}^{i,j}.
\label{h18}
\end{eqnarray}
Then $H_{i,j}|\Psi _{s}\rangle = 0$ and, accordingly, $H|\Psi _{s}\rangle =
0$.

Consequently, $\Psi _{s}$ is an exact ground-state wave function.

\begin{figure}[t]
\unitlength1cm
\begin{picture}(11,6)
\centerline{\psfig{file=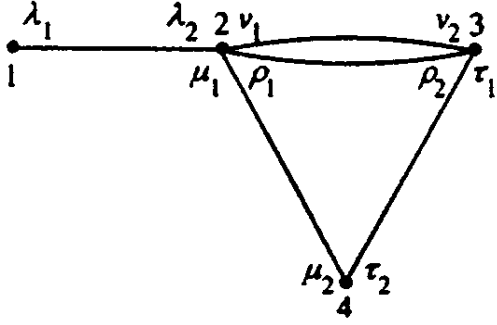,angle=0,width=8cm}}
\end{picture}
\caption[]{Example of a graph corresponding to the wave
function~(\ref{h19}).\label{jetp9}}
\end{figure}

We note that any two lattice sites can be joined by two, three, or more
bonds, because this does not contradict the principle of construction of
the wave function.  Moreover, the general principle of construction of the
wave function is valid not only for translationally symmetric lattices, but
for any graph in general.  As an example, let us consider the system shown
in Fig.~\ref{jetp9}.  The wave function of this system has the form
\begin{eqnarray}
\Psi _{s} = \Psi ^{\lambda _{1}}(1)\Psi ^{\lambda _{2}\mu _{1}\nu _{1}\rho
_{1}}(2)\Psi ^{\rho _{2}\nu _{2}\tau _{1}}(3)\Psi ^{\mu _{2}\tau _{2}}(4)
g_{\lambda _{1}\lambda _{2}}g_{\mu _{1}\mu _{2}}g_{\nu _{1}\nu _{2}}g_{\rho
_{1}\rho _{2}}g_{\tau _{1}\tau _{2}}
\label{h19}
\end{eqnarray}
and describes a system containing ten spins~1/2.

If the given lattice has dangling bonds (as occurs for systems with open
boundary conditions), the resulting wave function represents a spinor of
rank equal to the number of loose ends.  The ground state of this kind of
system is therefore $2^{l}$-fold degenerate, where {\it l} is the number of
loose ends.  For an open one-dimen\-sional chain, for example, the ground
state corresponds to four functions --- one singlet and three triplet
components.  For higher-dimen\-sional lattices this degeneracy depends on
the size of the lattice and increases exponentially as its boundaries grow.

\section{Electronic models}

In recent years there has been increasing interest in studying models where 
at least the ground state can be found exactly \cite{32,34}. The most popular 
method for the construction of exact ground state is so-called optimal ground 
state approach (OGS) \cite{34}. In the OSG method the ground state of the 
system is simultaneously the ground state of each local interaction. 
In this Section we propose new 1D and 2D models of interacting electrons with 
the exact ground state which are different from those constructed in the OSG 
method. The ground state wave function of our models is
expressed in terms of the two-particle {\it `singlet bond'} (SB) function
located on sites $i$ and $j$ of the lattice:
\begin{equation}
[i,j] = c_{i,\uparrow}^{+} c_{j,\downarrow}^{+}
- c_{i,\downarrow}^{+} c_{j,\uparrow}^{+}
+ x\,(c_{i,\uparrow}^{+} c_{i,\downarrow}^{+}
+ c_{j,\uparrow}^{+} c_{j,\downarrow}^{+}) \left| 0 \right \rangle ,
\label{k1}
\end{equation}
where $c_{i,\sigma}^{+}$, $c_{i,\sigma}$ are the Fermi operators and $x$ is an
arbitrary coefficient. The SB function is the generalization of the RVB
function \cite{38} including ionic states. The presence of the ionic states
is very important from the physical point of view because, as a rule, the bond 
functions contain definite amount of the ionic states as well.

It is known a set of 1D and 2D quantum spin models the exact ground state of
which can be represented in the RVB form \cite{42,39,44,40,jetp}. It is
natural to try to find electronic models with exact ground state at
half-filling formed by SB functions in the same manner as for above mentioned
spin models. The electronic models of these types include the correlated
hopping of electrons as well as the spin interactions and pair hopping terms.


{\it The model with dimerization}.

As the first example we consider the 1D electronic model with the two-fold
degenerate ground state in the form of the simple product of SB dimers, 
similarly to the ground state of the well-known spin-$\frac 12$ Majumdar-Ghosh
model \cite{39}. For the half-filling case the proposed ground state wave
functions are:
\begin{equation}
\Psi_1 = [1,2][3,4]...[N-1,N]
\label{k2} 
\end{equation}
and
\begin{equation}
\Psi_2 = [2,3][4,5]...[N,1]
\label{k3}
\end{equation}

In order to find the Hamiltonian for which the wave functions (\ref{k2})
and (\ref{k3}) are the exact ground state wave functions, we represent the 
Hamiltonian as a sum of local Hamiltonians $h_i$ defined on three neighboring
sites (periodic boundary conditions are supposed): 
\begin{equation}
H = \sum_{i=1}^N h_i
\label{k4}
\end{equation}

The basis of three-site local Hamiltonians $h_i$ consists of 64 states, while
only eight of them are present in $\Psi_1$ and $\Psi_2$. These 8 states are
\begin{equation}
[i,i+1]\,\varphi _{i+2}, \qquad \varphi _{i}\,[i+1,i+2] ,
\label{k5}
\end{equation}
where $\varphi _{i}$ is one of four possible electronic states of
$i$-th site: $\left| 0\right \rangle _i$, $\left|\uparrow\right\rangle _i$, 
$\left|\downarrow \right \rangle _i$, $\left| 2 \right \rangle _i$.

The local Hamiltonian $h_i$ for which all the functions (\ref{k5})
are the exact ground state wave functions can be written as the sum of the
projectors onto other 56 states $\left|\chi_k \right \rangle$
\begin{equation}
h_i = \sum_{k} \lambda _{k}  
\left|\chi_k \right \rangle  \left \langle \chi_k \right| ,
\label{k6}
\end{equation}
where $\lambda_{k}$ are arbitrary positive coefficients. This means that the 
wave functions $\Psi_1$ and $\Psi_2$ are the ground states of each local
Hamiltonian with zero energy. Hence, $\Psi_1$ and $\Psi_2$ are the `optimal'
ground state wave functions of the total Hamiltonian $H$ with zero energy,
similarly to the models in \cite{13,34,24,25}.
In general case, the local Hamiltonian $h_i$ is many-parametrical and depends
on parameters $\lambda _k$ and $x$. We consider one of the simplest forms of
$h_i$ including the correlated hopping of electrons of different types and
spin interactions between nearest- and next-nearest neighbor sites:
\begin{eqnarray}
h_i &=& 2 - x\,(t_{i,i+1}+t_{i+1,i+2}) \nonumber \\
&+&  (x^2-(1+x^2)(1-n_{i+1})^2)\;T_{i,i+2} \nonumber \\
&+& 8\frac{1-x^2}{3}({\bf S}_{i}\cdot {\bf S}_{i+1}+{\bf S}_{i+1}\cdot {\bf
S}_{i+2}+{\bf S}_{i}\cdot {\bf S}_{i+2}) ,
\label{k7} 
\end{eqnarray}
where
\begin{eqnarray}
T_{i,j} &=& \sum_{\sigma}(c_{i,\sigma}^{+}c_{j,\sigma}+c_{j,\sigma}^{+}c_{i,\sigma})
(1-n_{i,-\sigma}-n_{j,-\sigma}) , \nonumber \\
t_{i,j} &=& \sum_{\sigma}   (c_{i,\sigma}^{+}c_{j,\sigma}+
c_{j,\sigma}^{+}c_{i,\sigma})(n_{i,-\sigma}-n_{j,-\sigma})^2
\label{k8}
\end{eqnarray}
and ${\bf S}_i$ is the $SU(2)$ spin operator.

Each local Hamiltonian $h_i$ is a non-negatively defined operator at 
$|x|\le 1$. The following statements related to the Hamiltonian (\ref{k7}) 
are valid:

1. The functions (\ref{k2}) and (\ref{k3}) are the only two ground state
wave functions of the Hamiltonian (\ref{k7}) at $N_e = N$ ($N_e$ is the total
number of electrons).  They are not orthogonal, but their overlap is $\sim
e^{-N}$ at $N\gg 1$. 

2. The ground state energy $E_0(N_e/N)$ is a symmetrical function with respect
to the point $N_e/N=1$ and has a global minimum $E_0=0$ at $N_e/N=1$.

3. The translational symmetry of (\ref{k7}) is spontaneously broken in the
ground state leading to the dimerization:
\[
\left \langle |t_{i,i+1}-t_{i+1,i+2}| \right \rangle = 2
\]

The excited states of the model can not be calculated exactly but we expect
that there has to be a gap, because the ground state is formed by the
ultrashort-range SB functions. If it is the case, the function $E_0(N_e/N)$
has a cusp at $N_e/N = 1$.

Actually, this model is the fermion version of the  Majumdar -- Ghosh spin
model. Moreover, it reduces to the Majumdar -- Ghosh model at $x=0$ and in the 
subspace with $n_i=1$.

For $x=1$ the Hamiltonian (\ref{k7}) simplifies and takes the form:
\begin{equation}
H = -2 \sum_{j} t_{j,j+1} - \sum_{j} e^{i\pi n_{j+1}} T_{j,j+2}
\label{k9}
\end{equation}


{\it The 2D model}.

\begin{figure}[t]
\unitlength1cm
\begin{picture}(8,4)
\centerline{\psfig{file=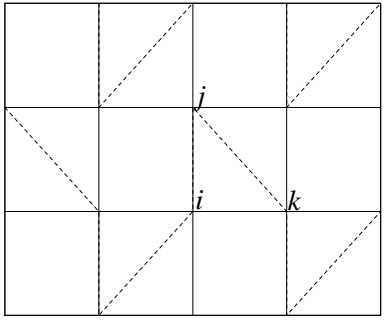,angle=-90,width=5cm}}
\end{picture}
\caption{ \label{shastry} The lattice of the Shastry -- Sutherland model}
\end{figure}

We can easily construct the 2D electronic model with the exact ground state
which is analogous to the Shastry -- Sutherland model \cite{44} 
(Fig.\ref{shastry}). The Hamiltonian of this model is:
\begin{equation}
H = \sum_{\{i,j,k\}} h_{i,j} + h_{i,k} + h^d_{j,k} ,
\label{k10}
\end{equation}
where the sum is over all triangles $\{i,j,k\}$, one of which is shown on
Fig.\ref{shastry}. So, each diagonal line belongs to the two different 
triangles. The local Hamiltonians $h^d_{j,k}$ acting on the diagonal of the 
triangle $\{i,j,k\}$, and $h_{i,j}$, $h_{i,k}$ have the form (for the sake 
of simplicity we put $x=1$)
\begin{eqnarray}
h^d_{j,k} &=& -2\;t_{j,k} + 4 \nonumber \\
h_{i,j}   &=& - t_{i,j} - e^{i\pi n_k}\;T_{i,j}  \nonumber \\
h_{i,k}   &=& - t_{i,k} - e^{i\pi n_j}\;T_{i,k}  
\label{k11}
\end{eqnarray}

It is easy to check that
\[
h^d_{j,k}\; \left| \varphi _{i}\;[j,k] \right\rangle
= (h_{i,j}+h_{i,k}) \;\left| \varphi _{i}\;[j,k]\right\rangle = 0   
\]

All other states of the local  Hamiltonian $h_{i,j}+h_{i,k}+h^d_{j,k}$ have
higher energies. Therefore, the ground state wave function in the half-filling
case is the product of the SB functions located on the diagonals shown by
dashed lines on Fig.\ref{shastry}. This model has non-degenerate singlet 
ground state with ultrashort-range correlations.


{\it The ladder model}.

Let us now consider electronic models with a more complicated ground state
including different configurations of short-range SB functions. The form
of these ground states is similar to that for spin models proposed in
\cite{40} and generalized in \cite{jetp}. In the 1D case our model describes
two-leg ladder model (Fig.\ref{lad}). Its ground state is a superposition of 
the SB functions where each pair of nearest neighbor rungs of the ladder is 
connected by one SB. 

The wave function of this ground state can be written in a form (\ref{g5}):
\begin{equation} 
\Psi _{s} = \psi ^{\lambda \mu }(1)g_{\mu \nu }\psi ^{\nu \rho }(2)g_{\rho
\kappa }\ldots \psi ^{\sigma \tau}(N)g_{\tau \lambda }
\label{k12}
\end{equation}
The functions $\psi ^{\lambda \mu }(i)$ describes $i$-th 
rung of the ladder
\begin{equation} 
\psi ^{\lambda \mu }(i) = c_1 \varphi _{2i-1}^{\lambda} \varphi _{2i}^{\mu } 
+ c_2 \varphi _{2i}^{\lambda} \varphi _{2i-1}^{\mu } 
\label{k13}
\end{equation}
with
\begin{equation} 
\varphi _{k}^{\lambda} = \left (
\begin{array}{c}
\left|\uparrow   \right \rangle _k \\
\left|\downarrow \right \rangle _k \\
\left| 2   \right \rangle _k\\
\left| 0   \right \rangle _k
\end{array} \right)
,\qquad
g_{\lambda \mu } = \left (
\begin{array}{cccc}
 0 & 1 & 0 & 0 \\
-1 & 0 & 0 & 0 \\
 0 & 0 & 0 & x \\
 0 & 0 & x & 0
\end{array}\right)
\label{k14}
\end{equation}

It is easy to see that
\[
g_{\lambda \mu }\varphi _{i}^{\lambda}\varphi _{j}^{\mu} = [i,j]
\]

Therefore, the function $\Psi_s$ is a singlet wave function depending on two
parameters $x$ and $c_1/c_2$. Actually, this form of $\Psi_s$ is equivalent to
the MP form with $4\times 4$ matrices $A_{\lambda \nu }(i) =
g_{\lambda\mu}\psi ^{\mu\nu}(i)$. Moreover, at $x=0$ and $c_1/c_2=-1$ the
function $\Psi_s$ reduces to the wave function of the well-known AKLT 
spin-$1$ model.

In order to find the Hamiltonian for which the wave function (\ref{k12}) is
the exact ground state wave function, it is necessary to consider what states
are present on the two nearest rungs in the $\Psi _s$. It turns out that there
are only 16 states from the total 256 ones in the product $\psi ^{\lambda \mu
}(i)g_{\mu \nu }\psi ^{\nu \rho }(i+1)$. The local Hamiltonian $h_i$ acting on
two nearest rungs $i$ and $i+1$ can be written in the form of (\ref{k6}) with
the projectors onto the 240 missing states. The total Hamiltonian is the sum
of local ones (\ref{k4}). The explicit form of this Hamiltonian is very
cumbersome and, therefore, it is not given here.

The correlation functions in the ground state (\ref{k12}) can be calculated
exactly in the same manner as it was done for spin models \cite{40}. It
can be shown that all of correlations exponentially decay in the ground state.
We expect also that this model has a gap.

This method of construction of the exact ground state can be generalized also
to 2D and 3D lattices, as it was done in Sec.IV. Following \cite{jetp}, 
one can rigorously prove that the ground state of these models is always a 
non-degenerate singlet.

{\it 1D models with the giant spiral order}.

There is one more spin-$\frac 12$ model with an exact ground state of the RVB
type \cite{42}. This is the model (\ref{f1}) describing the F-AF
transition point. The exact singlet ground state can be expressed by the 
combinations of the RVB functions $(i,j)$ distributed uniformly over the  
lattice points (\ref{a8}). The analog of the wave function (\ref{a8}) in the 
SB terms is:
\begin{equation}
\Psi _0 =\sum_{i<j\ldots} (-1)^P\;[i,j][k,l][m,n]\ldots  ,
\label{m1}
\end{equation}
where $P=(i,j,k,l,\ldots)$ is the permutation of numbers $(1,2,\ldots N)$.
It is interesting to note that the singlet wave function (\ref{m1}) can be
also written in the MP form but with an infinite size matrices \cite{ele}.

In order to find the Hamiltonian for which the wave function (\ref{m1}) 
is the exact ground state wave function, let us consider what 
states are present on the two nearest sites in the $\Psi _0$.
It turns out \cite{ele} that there are only 9 states from the total 16 ones
in (\ref{m1}). They are
\begin{eqnarray}
&&\left|\uparrow \uparrow \right \rangle ,\qquad
\left|\downarrow \downarrow  \right \rangle ,\qquad
\left|\uparrow \downarrow + \downarrow \uparrow  \right \rangle ,
\qquad \left|20-02 \right \rangle , 
\nonumber \\
&&\left|\uparrow \downarrow - \downarrow \uparrow  \right \rangle
+ x \: \left|20+02 \right \rangle ,\qquad
\left|\uparrow 0 - 0\uparrow \right \rangle ,\nonumber \\
&&\left|\uparrow 2 - 2\uparrow \right \rangle ,\qquad
\left|\downarrow 0 - 0\downarrow \right \rangle ,\qquad
\left|\downarrow 2 - 2\downarrow \right \rangle 
\label{m2}
\end{eqnarray}

The local Hamiltonian $h_{i,i+1}$ can be written as the sum of the projectors 
onto the 7 missing states (\ref{k6}). 
At $|x|>1$ the most simple form of this total Hamiltonian is:
\begin{eqnarray}
H &=& \sum_{i=1}^{N} \left(
T_{i,i+1} - \frac{2}{x} t_{i,i+1}
-4\:{\bf S}_{i}\cdot {\bf S}_{i+1} \right.
\nonumber \\
&+& \left. \frac{4}{x^2}{\bf \eta}_{i}\cdot {\bf \eta}_{i+1}
+ 4\,\frac{x^2-3}{x^2}\eta_{i}^z\eta_{i+1}^z + 1 \right)
\label{m3}
\end{eqnarray}
We use here ${\bf \eta}$ operators:
\[
\eta_i^{+}=c_{i,\downarrow }^{+}\:c_{i,\uparrow }^{+}, \qquad
\eta_i^{-}=c_{i,\uparrow }\:c_{i,\downarrow }, \qquad
\eta_i^{z}=\frac {1-n_i}{2} , 
\]
which form another $SU(2)$ algebra \cite{36,37}, and 
${\bf \eta}_{1}\cdot {\bf \eta}_{2}$ is a scalar product of pseudo-spins
${\bf \eta}_{1}$ and ${\bf \eta}_{2}$. We note that this Hamiltonian commutes
with ${\bf S}^2$, but does not commute with ${\bf \eta}^2$.
It can be proved \cite{ele} that only three multiplets are the ground states of
(\ref{m3}): the singlet state (\ref{m1}), the trivial ferromagnetic state
$S=N/2$ and the state with $S=N/2-1$.

The norm and the correlators of the electronic model (\ref{m3}) in the
singlet ground state are exactly calculated \cite{ele}. For example, the norm
of (\ref{m1}) is:
\[
\langle \Psi _0| \Psi _0\rangle  = \left. \frac {d^{N}}{d\xi ^{N}} 
\left( 2\frac{1+\cosh(x \xi )}{\cos^2(\frac{\xi}{2})}  \right)  \right|_{\xi=0}
\]

The correlators at $N\gg 1$ are
\begin{eqnarray}
\langle \eta_i^z \eta_{i+l}^z \rangle &=& O\left(\frac {1}{N^2}\right) 
,\qquad
\langle \eta_i^- \eta_{i+l}^+ \rangle = O\left(\frac {1}{N^2}\right) ,
\nonumber \\
\langle c_{i,\sigma}^{+}\:c_{i+l,\sigma}\rangle &=& O\left(\frac 1N\right) ,
\qquad
\left\langle {\bf S}_{i}{\bf S}_{i+l}\right\rangle =
\frac{1}{4}\cos \left( \frac{2\pi l}{N}\right)
\label{m4} 
\end{eqnarray}

The correlator $\langle \eta_i^{+} \eta_{i+l}^{-} \rangle $ which determines
the off-diagonal long-range order (ODLRO) \cite{35} vanishes in the
thermodynamic limit. At the same time the spin-spin correlations have a spiral 
form, and the period of the spiral equals to the system size as in the spin 
model (\ref{f1}).


Another electronic model can be obtained by making the canonical
transformation $c^{+}_{i,\uparrow}\to c^{+}_{i,\uparrow}$ and
$c^{+}_{i,\downarrow}\to c_{i,\downarrow}$. As a result of this
transformation, the SB function (\ref{k1}) becomes:
\begin{equation}
\{i,j\} = c_{i,\uparrow}^{+} c_{i,\downarrow}^{+}
- c_{j,\uparrow}^{+} c_{j,\downarrow}^{+}
- x\,(c_{i,\uparrow}^{+} c_{j,\downarrow}^{+}
+ c_{i,\downarrow}^{+} c_{j,\uparrow}^{+} ) \left| 0 \right \rangle ,
\label{m5}
\end{equation}
and the wave function (\ref{m1}) changes to
\begin{equation}
\Psi _0 =\sum_{i<j\ldots} \{i,j\}\{k,l\}\{m,n\}\ldots
\label{m6}
\end{equation}

The function (\ref{m6}) for $|x|>1$ is the exact ground state wave function
of the transformed Hamiltonian:
\begin{eqnarray}
H &=& \sum_{i=1}^{N} \left(
-T_{i,i+1} - \frac{2}{x} t_{i,i+1}^{\prime}
-4{\bf \eta}_{i}\cdot {\bf \eta}_{i+1} \right.
\nonumber \\
&+& \left.\frac{4}{x^2}{\bf S}_{i}\cdot {\bf S}_{i+1}
+4\,\frac{x^2-3}{x^2} S_{i}^z S_{i+1}^z +1 \right) ,
\label{m7}
\end{eqnarray}
where
\[
t_{i,i+1}^{\prime} = \sum_{\sigma} \sigma(c_{i,\sigma}^{+}c_{i+1,\sigma}+
c_{i+1,\sigma}^{+}c_{i,\sigma})(n_{i,-\sigma}-n_{i+1,-\sigma})^2
\]
This Hamiltonian commutes with ${\bf \eta}^2$ but does not commute with ${\bf
S}^2$. Therefore, the eigenfunctions of the Hamiltonian (\ref{m7}) can be
described by quantum numbers ${\bf\eta}$ and $\eta^z$. For the cyclic model 
the states with three different values of $\eta$ have zero 
energy \cite{ele} [as it was for the model (\ref{m3})]. They include
one state with $\eta =0$ (\ref{m6}), all states with $\eta =N/2$:
\begin{equation}
\Psi_{N/2,\;\eta^z} = (\eta^+)^{N/2-\eta^z} \left|0 \right \rangle ,
\label{m8}
\end{equation}
and the states with $\eta =N/2-1$.
Therefore, for the case of one electron per site ($\eta^z=0$), the ground 
state of the model (\ref{m7}) is three-fold degenerate.

The correlation functions in the ground states with $\eta =N/2$ and $\eta
=N/2-1$ for the half-filling case coincide with each other and at $N\gg 1$
they are:
\begin{eqnarray}
\langle c_{i,\sigma}^{+}\:c_{i+l,\sigma}\rangle = O\left(\frac 1N\right) , 
\qquad
\langle {\bf S}_{i}{\bf S}_{i+l} \rangle = O\left(\frac {1}{N^2}\right) ,
\nonumber \\
\left\langle \eta_i^z \eta_{i+l}^z \right\rangle = O\left(\frac 1N\right) ,
\qquad
\left\langle \eta_i^- \eta_{i+l}^+ \right\rangle = \frac{1}{4} + 
O\left(\frac 1N\right) 
\label{m9} 
\end{eqnarray}
The existence of the ODLRO immediately follows from the latter equations.
The correlation functions in the ground state (\ref{m6}) have similar 
forms as in Eqs.(\ref{m4}):
\begin{eqnarray}
&&\langle c_{i,\sigma}^{+}\:c_{i+l,\sigma}\rangle = O\left(\frac {1}{N}\right),
\qquad 
\langle {\bf S}_{i}{\bf S}_{i+l} \rangle = O\left(\frac {1}{N^2}\right),
\nonumber \\
&&\left\langle \eta_i^- \eta_{i+l}^+ \right\rangle =
2 \left\langle \eta_i^z \eta_{i+l}^z \right\rangle =
\frac{1}{6}\cos \left( \frac{2\pi l}{N}\right) 
\label{m10} 
\end{eqnarray}
The giant spiral ordering in the last equation implies the existence of the
ODLRO and, therefore, the superconductivity \cite{35} in the ground state 
(\ref{m6}). 

Similarly to the original spin model (\ref{f1}) \cite{prb,ko} the last two
electronic models (\ref{m3}),(\ref{m7}) describe the transition points
on the phase diagram between the phases with and without a long-range order
[ferromagnetic for the model (\ref{m3}) and off-diagonal for the model
(\ref{m7})]. Therefore, we suggest the formation of the ground state with
giant spiral order (ferromagnetic or off-diagonal) as a probable scenario of
the subsequent destruction of the ferromagnetism and superconductivity.

\section{Conclusion}

We have considered the class of the 1D and 2D spin and electronic models
with the exact ground states.

The one of these models is the spin-$\frac 12$ ladder with competing 
interactions of the ferro- and antiferromagnetic types at the F-AF 
transition line. The exact singlet ground state wave function on this 
line is found in the special form expressed in terms of auxiliary 
Bose-operators. The spin correlators in the singlet state show 
double-spiral ordering with the period of spirals equal to the system size.

In general case the proposed form of the wave function corresponds to the MP
form but with matrices of infinite size. However, for special values of
parameters of the model it can be reduced to the standard MP form. In
particular, we consider spin-$\frac 12$ ladder with nondegenerate 
antiferromagnetic ground state for which the ground state wave function is 
the MP one with $2\times 2$ matrices. This model has some properties of 
1D AKLT model and  reduces to it in definite limiting case.

The ground state wave function of the spin ladders can be represented in an
alternative form as a product of second-rank spinors associated with the
lattice sites and the metric spinors corresponding to bonds between nearest
neighbor sites. Two-dimensional spin-$\frac 12$ model is constructed with exact
ground state wave function of this type. The ground state of this model is a
nondegenerate  singlet with exponentially decay of spin correlators. We
believe the model has a gap in the spectrum of excitations.

We propose new models of interacting electrons with the exact ground state
formed by the singlet bond functions in the same manner as for some spin
models.  In particular, we considered the models describing the boundary
points on the phase diagram between the phases with and without long-range
order (ferromagnetic or off-diagonal).

In conclusion we note that the construction of considered models is based on
following property. Their Hamiltonians are the sums of the Hamiltonians that
are local and non-commuting with each other. In the same time the ground
state wave function of the total Hamiltonian is the ground state one for
each for them. It is clear that these models are rather special.
Nevertheless, the study of them is useful for understanding properties of
the real frustrated spin systems and strongly correlated electronic models.


This work was supported by the Russian Foundation for Basic 
Research (grants no.00-03-32981, 99-03-3280 and no.00-15-97334).

\end{document}